\documentclass[12pt, preprint]{aastex}

\newcommand{\be}{\begin{enumerate}}
\newcommand{\ee}{\end{enumerate}}
\newcommand{\dg}{$^{\circ}$}

\shorttitle{Barstrength and Circumnuclear Dust}
\shortauthors{Peeples \& Martini}

\begin{document}

\title{The Connection Between Barstrength and Circumnuclear Dust Structure\altaffilmark{1}}

\author{Molly S.\ Peeples, 
Paul Martini \affil{Department of Astronomy, Ohio State University,
140 W.\ 18th Ave., Columbus,~OH~43210, molly@astronomy.ohio-state.edu,
martini@astronomy.ohio-state.edu}}

\altaffiltext{1}{Based on observations with the NASA/ESA {\it Hubble Space
Telescope} obtained at the the Space Telescope Science Institute, which is
operated by the Association of Universities for Research in Astronomy,
Incorporated, under NASA contract NAS5-26555.}

\begin{abstract}
We present a comparison of barstrength $Q_b$ and circumnuclear dust
morphology for 75 galaxies in order to investigate how bars affect the
centers of galaxies.  We trace the circumnuclear dust morphology and
amount of dust structure with structure maps generated from
visible-wavelength HST data, finding that tightly wound nuclear dust
spirals are primarily found in weakly barred galaxies.  While strongly
barred galaxies sometimes exhibit grand design structure within the
central 10\% of $D_{25}$, this structure rarely extends to within $\sim
10$~pc of the galaxy nucleus.  In some galaxies, these spiral arms
terminate at a circumnuclear starburst ring.  Galaxies with
circumnuclear rings are generally more strongly barred than galaxies
lacking rings.  Within these rings, the dust structure is fairly smooth
and usually in the form of a loosely wound spiral.  These data
demonstrate that multiple nuclear morphologies are possible in the most
strongly barred galaxies: chaotic central dust structure inconsistent
with a coherent nuclear spiral, a grand design spiral that loses
coherence before reaching the nucleus, or a grand design spiral that
ends in a circumnuclear ring.  These observations may indicate that not
all strong bars are equally efficient at fueling material to the centers
of their host galaxies.  Finally, we investigate the longstanding
hypothesis that SB(s) galaxies have weak bars and SB(r) galaxies have
strong bars, finding the opposite to be the case: namely, SB(r) galaxies
are less strongly barred and have less dust structure than SB(s)
galaxies.  In general, more strongly barred galaxies tend to have higher
nuclear dust contrast.
\end{abstract}


\section{Introduction}\label{sec:intro}
Most spiral galaxies have large-scale bars, and any complete picture of
galaxies must include the impact of bars on their evolution.  Bars are
important because they are an obvious mechanism for funneling gas and
dust to the centers of galaxies: gas clouds orbiting in the disk lose
angular momentum as they encounter the bar, thereby sinking towards the
galaxy's center \citep{binney87, athanassoula92, piner95, regan97,
maciejewski02}.  Bars are therefore expected to play a major role in
circumnuclear star formation, bulge growth, and the fueling of the
central, supermassive black hole.

Bar-driven radial gas inflow should drive large quantities of gas and
dust into the central regions of galaxies, thereby increasing the
central gas concentration of barred galaxies relative to unbarred
galaxies.  Observations by \citet{sakamoto99} and \citet{sheth05}
support this view, finding higher concentrations of molecular gas in the
central kiloparsec of barred galaxies than in unbarred galaxies. As
higher gas density is empirically correlated with higher star formation
rates \citep[e.g.,][]{kennicutt98}, there should be a correlation
between bar-driven inflow and central star formation.  Observational
evidence for enhanced nuclear or circumnuclear star formation in barred
galaxies provides broad support for this picture \citep{ho97c, maoz01,
knapen06}.

It has long been speculated that this same radial gas inflow could
provide active galactic nuclei (AGN) with fuel \citep{simkin80,
schwarz81}.  There is so far no evidence, though, that bars (large-scale
or nuclear) are the primary fuel source for AGN: not all active galaxies
are barred, not all barred galaxies are active, and, in fact,
well-matched samples of active and inactive galaxies have the same bar
fraction \citep{ho97c, mulchaey97a, laine02}.  \citeauthor{ho97c}
speculate that this is because radially-transported gas is prevented
from reaching the nucleus.

If bars are indeed fueling galaxies' circumnuclear regions, then this
should be reflected in the morphology of the cold circumnuclear
interstellar medium (ISM).  \citet{martini03a} establish a nuclear dust
morphology classification system with four spiral classes (grand design,
tightly wound, loosely wound, and chaotic spiral) and two non-spiral
classes (chaotic and no structure) to quantify differences between
various galaxy types.  \citet{martini03b} find no differences in the
nuclear dust structure of active and inactive galaxies.  This study does
find evidence that nuclear grand design spirals are primarily found in
barred galaxies and that these grand design spirals connect to large
scale dust lanes in barred galaxies.  Likewise, \citeauthor{martini03b}
primarily find tightly wound nuclear spirals in unbarred galaxies.

All of these studies of the role of bars in fueling central star
formation or black holes simply compare ``barred'' versus ``unbarred''
galaxies.  This discretization glosses over, and over simplifies, the
long-known continuum of barstrengths \citep{devaucouleurs59}. One early
way of quantifying the strength of a bar is the deprojected bar
ellipticity $\epsilon_b$ \citep{martin95}.  The disadvantage of using
the bar ellipticity is that defining the bar---i.e., its exact
dimensions---is a somewhat subjective process \citep{buta01}.
\citet{abraham00} introduce the parameter $f_{\mbox{\scriptsize bar}}$,
which is a function of the bar axis ratio, and therefore susceptible to
the same systematics as $\epsilon_b$.  \citeauthor{buta01}, by expanding
on the method of \citet{combes81}, circumvent this problem by using a
force ratio, which they call $Q_b$.  Assuming that the light traces the
underlying mass distribution, they use near-infrared images to calculate
a map of a galaxy's potential.  From these potential maps, and with some
assumptions about the characteristic scale-height, they then calculate
the tangential force $F_T$ and mean (axisymmetric) radial force $\langle
F_R\rangle$.  They define the ratio map $Q_T$ as
\begin{equation}\label{eqn:Qt}
Q_T(i,j) = \frac{F_T(i,j)}{\langle F_R(i,j)\rangle}.
\end{equation}
This ratio map has the property that in each of four quadrants $Q_T$
reaches a local extremum.  Letting $Q_{T}^{\mbox{\scriptsize max},\,k}$
be the absolute value of such an extremum in the $k$th quadrant, the
barstrength $Q_b$ is defined as
\begin{equation}\label{eqn:Qb}
Q_b = \frac{1}{4} \sum\limits^{4}_{k=1}  Q_{T}^{\mbox{\scriptsize max},\,k}.
\end{equation}
The cited uncertainties on $Q_b$ are a measure of how much
$Q_{T}^{\mbox{\scriptsize max}}$ varies by quadrant.  From a comparison
with the three barstrength classes defined by \citeauthor{devaucouleurs59}
(unbarred SA, weakly barred SAB, and strongly barred SB),
\citeauthor{buta01} find a reasonable correlation between the RC3 bar
classification and $Q_b$, although with substantial scatter.
In particular, 
SA galaxies have $Q_b \lesssim 0.1$, SAB galaxies have $0.05 \lesssim
Q_b \lesssim 0.2$, and SB galaxies have $Q_b \gtrsim 0.15$.  The
most strongly barred galaxies have $Q_b \approx 0.6$ \citep{buta01}.
While we adopt $Q_b$ as the best available measure of barstrength, we
note that \citet{athanassoula92} shows that both the quadrupole moment
(strength) and the pattern speed play an important role in how
effectively the bar drives mass towards the center of the galaxy.
Pattern speed, however, is much more difficult to measure than
parameters that can be calculated from photometric data.

\citet{martini04} used $Q_b$ and the nuclear classifications of
\citet{martini03a} to compare the barstrengths and nuclear dust
morphologies of 48 galaxies with archival Hubble Space Telescope (HST)
data.  He found that grand design nuclear spirals are primarily found in
strongly barred galaxies, while axisymmetric tightly wound nuclear
spirals are exclusively found in galaxies with $Q_b < 0.1$.  With a
larger sample and reconsideration of the circumnuclear dust morphology
classification at the smallest scales, we examine how nuclear dust
structure varies with barstrength.  We identify 75 galaxies, described
in \S\ref{sec:data}, with a measured barstrength $Q_b$ from the
literature.  To analyze the dust morphology, we create ``structure
maps'' from archival HST data using the technique of \citet{pogge02} as
discussed in \S\ref{sec:smap}.  We then classify the galaxies according
to a refined version of the classification scheme proposed in
\citet{martini03a}, as described in \S\ref{sec:nuc}.  The main
difference between the classification scheme used here and that of
\cite{martini03a} is a stronger focus on the central-most regions of the
galaxy.  We discuss our results in \S\ref{sec:results}.

\section{Data}\label{sec:data}
We began our study with a data set of 127 galaxies with both a measured
barstrength (primarily from the Ohio State University Bright Galaxy
Survey (OSUBGS) by \citealt{laurikainen04}) and archival HST data.  The
HST images were obtained with WFPC2, with the exception of three images
from the Advanced Camera for Surveys (ACS), and most employed either the
F606W or F814W filters.  The galaxies studied by the OSUBGS were
selected to be a representative sample of the nearby universe;
specifically, based on the RC3 T-type distribution ($0 \le$~T~$\le 9$,
or S0/a to Sm) to ensure there is no morphological selection bias
\citep{eskridge02}.
\clearpage
\begin{center}
\begin{deluxetable}{lclccccclll}
\tabletypesize{\scriptsize}
\tablecaption{Data
\label{tbl:data} }
\tablecolumns{11}
\tablehead{
\colhead{Galaxy} &
\colhead{Nuclear} &
\colhead{$Q_b$} & 
\colhead{$\sigma_{\mbox{\scriptsize sm}}$} &
\colhead{Detector} &
\colhead{Filter} & 
\colhead{HST} &  
\colhead{T} &
\colhead{$D$} & 
\colhead{$r_c$} & 
\colhead{Notes} \\ 
\colhead{} &
\colhead{Class} &
\colhead{} & 
\colhead{$(\times 10^{-2})$} &
\colhead{} &
\colhead{} & 
\colhead{Prop.\ ID} &  
\colhead{} &
\colhead{(Mpc)} & 
\colhead{(pc)} & 
\colhead{} \\
\colhead{(1)} &
\colhead{(2)} &
\colhead{(3)} & 
\colhead{(4)} &
\colhead{(5)} &
\colhead{(6)} & 
\colhead{(7)} &
\colhead{(8)} &
\colhead{(9)} &
\colhead{(10)} & 
\colhead{(11)}
}
\startdata
NGC289    &  LW   &    0.212 $\pm$ 0.003     &  1.280 $\pm$ 0.030 &	 PC1  &        F606W &	   6359 &   4	&  19.4 & 221\tablenotemark{d}  &    SB(rs) \\
NGC488    &  LW   &    0.032 $\pm$ 0.003     &  0.951 $\pm$ 0.015 &	 PC1  &        F606W &	   6359 &   3	&  29.3 & 447  & \\
NGC613    &  CS   &    0.401 $\pm$ 0.045     &  4.469 $\pm$ 0.017 &	 WF3  &        F606W &	   9042 &   4	&  17.5 & 280  &    SB(rs) \\
NGC864    &  CS   &    0.360 $\pm$ 0.037     &  3.726 $\pm$ 0.064 &	 PC1  &        F606W &	   8597 &   5	&  20 &   177  & LGD\\
NGC972    &  C    &    0.220 $\pm$ 0.060\tablenotemark{a}  &  3.948 $\pm$ 0.060  &	  PC1  &       F606W &	   6359 &   2	&  21.4 & 206 & \\
NGC1068   &  C    &    0.165 $\pm$ 0.010     &  4.204 $\pm$ 0.072 &	 PC1  &        F606W &	   8597 &   3	&  14.4 & 134\tablenotemark{d}  & \\
NGC1073   &  C    &    0.607 $\pm$ 0.013     &  1.839 $\pm$ 0.013 &	 WF3  &        F606W &	   9042 &   5	&  15.2 & 217  &    SB(rs) \\
NGC1097   &  LW   &    0.279 $\pm$ 0.048     &  2.416 $\pm$ 0.025 &	 HRC  &        F814W &	   9788 &   3	&  14.5 & 394  &    SB(s), LGD \\
NGC1241   &  LW   &    0.251 $\pm$ 0.028     &  3.383 $\pm$ 0.037 &	 PC1  &        F606W &	   5479 &   3	&  46.6 & 382  &    SB(rs), LGD \\
NGC1300   &  LW   &    0.537 $\pm$ 0.011     &  2.362 $\pm$ 0.025 & PC1  &        F606W &	   8597 &   4	&  18.8 & 311\tablenotemark{d} &    SB(rs), LGD \\
NGC1317   &  CS   &    0.085 $\pm$ 0.007     &  1.910 $\pm$ 0.023 &	 PC1  &        F606W &	   5446 &   1	&  16.9 & 135  &    LGD \\
NGC1326   &  CS   &    0.160 $\pm$ 0.020\tablenotemark{b}  &  2.506 $\pm$ 0.019 &	 PC1  &        F555W &	   6496 &   -1  &  16.9 & 191 & LGD\\
NGC1350   &  C    &    0.243 $\pm$ 0.039     &  0.863 $\pm$ 0.008 &	 PC1  &        F606W &	   5446 &   1.8 &  16.9 & 258  &    SB(r) \\
NGC1365   &  CS   &    0.400 $\pm$ 0.110\tablenotemark{a}  &  3.892 $\pm$ 0.008  &	  PC1  &       F606W &	   8597 &   3	&  16.9 & 552 &    SB(s), LGD \\
NGC1385   &  C    &    0.319 $\pm$ 0.030     &  2.298 $\pm$ 0.024 &	 PC1  &        F606W &	   6359 &   6	&  17.5 & 172  &    SB(s) \\
NGC1398   &  N   &    0.202 $\pm$ 0.011     &  0.552 $\pm$ 0.001 &	 PC1  &        F606W &	   8597 &   2	&  16.1 & 332  &    SB(r) \\
NGC1433   &  GD   &    0.370 $\pm$ 0.060\tablenotemark{a}  &  1.772 $\pm$ 0.028  &	  WF3  &       F814W &	   9042 &   2	&  11.6 & 218 &    SB(r) \\
NGC1530   &  TW   &    0.610 $\pm$ 0.160\tablenotemark{a}  &  3.965 $\pm$ 0.005  &	  PC1  &       F606W &	   8597 &   3	&  36.6 & 487 &    SB(rs), LGD \\
NGC1637   &  C    &    0.202 $\pm$ 0.014     &  2.054 $\pm$ 0.061 &	 WF4  &        F555W &	   9155 &   5	&  8.9 &  91\tablenotemark{d}  &     LGD \\
NGC1808   &  LW   &    0.274 $\pm$ 0.001     &  5.326 $\pm$ 0.046 &	 PC1  &        F675W &	   6872 &   1	&  10.8 & 203  & \\
NGC2196   &  C    &    0.070 $\pm$ 0.005     &  0.858 $\pm$ 0.017 &	 PC1  &        F606W &	   6359 &   1	&  28.8 & 236  & \\
NGC2442   &  CS   &    0.669 $\pm$ 0.428     &  2.559 $\pm$ 0.031 &	 WF3  &        F606W &	   9042 &   3.7 &  17.1 & 273  & \\
NGC2655   &  CS   &    0.128 $\pm$ 0.004     &  1.822 $\pm$ 0.003 &	 PC1  &        F547M &	   5419 &   0	&  24.4 & 348  & \\
NGC2775   &  N   &    0.050 $\pm$ 0.010     &  0.579 $\pm$ 0.009 &	 PC1  &        F606W &	   6359 &   2	&  17 &   211  & \\
NGC2964   &  C    &    0.310 $\pm$ 0.003     &  6.463 $\pm$ 0.162 &	 PC1  &        F606W &	   6359 &   4	&  21.9 & 149\tablenotemark{d}  &    LGD \\
NGC2985   &  TW   &    0.056 $\pm$ 0.001     &  1.608 $\pm$ 0.033 &	 WF2  &        F606W &	   5479 &   2	&  22.4 & 298  & \\
NGC2997   &  TW   &    0.060 $\pm$ 0.020\tablenotemark{b}  &  3.498 $\pm$ 0.067 &	 HRC  &        F555W &	   9989 &   5	&  13.8 & 358 & \\
NGC3054   &  C    &    0.170 $\pm$ 0.020\tablenotemark{b}  &  1.697 $\pm$ 0.013 &	 PC1  &        F606W &	   6359 &   3	&  26.5 & 207 & \\
NGC3077   &  C    &    0.119 $\pm$ 0.016     &  6.253 $\pm$ 0.093 &	 WF3  &        F814W &	   9144 &   90  &  2.1 &  33  & \\
NGC3081   &  GD   &    0.170 $\pm$ 0.020\tablenotemark{b}  &  2.501 $\pm$ 0.023 &	 PC1  &        F606W &	   5479 &   0	&  32.5 & 198 &    LGD \\
NGC3169   &  C    &    0.090 $\pm$ 0.005     &  3.254 $\pm$ 0.066 &	 PC1  &        F547M &	   5149 &   1	&  19.7 & 250  & \\
NGC3227   &  C    &    0.158 $\pm$ 0.021     &  4.615 $\pm$ 0.109 &	 PC1  &        F606W &	   5479 &   1	&  20.6 & 181\tablenotemark{d}  & \\
NGC3310   &  C    &    0.060 $\pm$ 0.010\tablenotemark{c}  &  1.575 $\pm$ 0.007  &	  PC1  &       F814W &	   6639 &   4	&  18.7 & 168 & \\
NGC3338   &  TW   &    0.083 $\pm$ 0.005     &  1.318 $\pm$ 0.011 &	 WF3  &        F606W &	   9042 &   5	&  22.8 & 391  & \\
NGC3359   &  C    &    0.460 $\pm$ 0.050\tablenotemark{c}  &  1.914 $\pm$ 0.002  &	  WF3  &       F606W &	   9042 &   5	&  19.2 & 405 &    SB(rs) \\
NGC3486   &  GD   &    0.108 $\pm$ 0.002     &  0.968 $\pm$ 0.006 &	 PC1  &        F606W &	   8597 &   5	&  12.3 & 165\tablenotemark{d}  & \\
NGC3504   &  LW   &    0.288 $\pm$ 0.030     &  3.862 $\pm$ 0.125 &	 PC1  &        F606W &	   5479 &   2	&  26.5 & 207  & \\
NGC3898   &  C    &    0.047 $\pm$ 0.000     &  0.877 $\pm$ 0.009 &	 PC1  &        F606W &	   6359 &   2	&  21.9 & 278  & \\
NGC4027   &  C    &    0.623 $\pm$ 0.008     &  2.143 $\pm$ 0.021 &	 PC1  &        F814W &	   8599 &   8	&  25.6 & 175\tablenotemark{d}  &    SB(s) \\
NGC4030   &  TW   &    0.060 $\pm$ 0.013     &  2.193 $\pm$ 0.026 &	 PC1  &        F606W &	   6359 &   4	&  25.9 & 314  & \\
NGC4051   &  CS   &    0.280 $\pm$ 0.008     &  2.873 $\pm$ 0.082 &	 PC1  &        F606W &	   5479 &   4	&  17 &   260  & \\
NGC4138   &  C    &    0.046 $\pm$ 0.007     &  2.829 $\pm$ 0.008 &	 PC1  &        F547M &	   6837 &   -1  &  17 &   127  & \\
NGC4254   &  LW   &    0.122 $\pm$ 0.029     &  2.640 $\pm$ 0.029 &	 PC1  &        F606W &	   8597 &   5	&  16.8 & 262  & \\
NGC4303   &  LW   &    0.259 $\pm$ 0.044     &  3.928 $\pm$ 0.062 &	 HRC  &        F555W &	   9776 &   4	&  15.2 & 218\tablenotemark{d}  &    LGD \\
NGC4314   &  LW   &    0.442 $\pm$ 0.024     &  1.390 $\pm$ 0.029 &	 PC1  &        F606W &	   8597 &   1	&  9.7 &  118  &    SB(rs), LGD \\
NGC4321   &  LW   &    0.183 $\pm$ 0.027     &  2.123 $\pm$ 0.045 &	 PC1  &        F702W &	   5195 &   4	&  16.8 & 362  &    LGD \\
NGC4414   &  LW   &    0.149 $\pm$ 0.003     &  1.857 $\pm$ 0.032 &	 PC1  &        F606W &	   8597 &   5	&  9.7 &  118  & \\
NGC4450   &  C    &    0.131 $\pm$ 0.011     &  0.634 $\pm$ 0.005 &	 PC1  &        F814W &	   5375 &   2	&  16.8 & 256  & \\
NGC4501   &  LW   &    0.072 $\pm$ 0.026     &  1.834 $\pm$ 0.023 &	 PC1  &        F606W &	   6359 &   3	&  16.8 & 276\tablenotemark{d}  & \\
NGC4504   &  C    &    0.136 $\pm$ 0.019     &  1.723 $\pm$ 0.016 &	 WF3  &        F814W &	   9042 &   6	&  19.5 & 248  & \\
NGC4579   &  CS   &    0.197 $\pm$ 0.020     &  1.713 $\pm$ 0.032 &	 PC1  &        F547M &	   6436 &   3	&  16.8 & 288  & \\
NGC4593   &  LW   &    0.309 $\pm$ 0.020     &  1.963 $\pm$ 0.069 &	 PC1  &        F606W &	   5479 &   3	&  39.5 & 447  &    SB(rs), LGD \\
NGC4651   &  GD   &    0.120 $\pm$ 0.046     &  1.329 $\pm$ 0.022 &	 PC1  &        F555W &	   5375 &   5	&  16.8 & 195  & \\
NGC4698   &  N   &    0.084 $\pm$ 0.040     &  0.700 $\pm$ 0.006 &	 PC1  &        F606W &	   6359 &   2	&  16.8 & 195  & \\
NGC4736   &  LW   &    0.048 $\pm$ 0.004     &  1.684 $\pm$ 0.003 &	 PC1  &        F555W &	   5741 &   2	&  4.3 &  140  & \\
NGC4939   &  C    &    0.128 $\pm$ 0.052     &  2.297 $\pm$ 0.008 &	 PC1  &        F606W &	   5479 &   4	&  17.3 & 113  & \\
NGC4941   &  CS   &    0.056 $\pm$ 0.008     &  1.582 $\pm$ 0.017 &	 PC1  &        F606W &	   8597 &   2	&  44.3 & 531\tablenotemark{d}  &    LGD \\
NGC5054   &  CS   &    0.090 $\pm$ 0.023     &  1.790 $\pm$ 0.032 &	 PC1  &        F606W &	   8597 &   4	&  9.7 &  102  & \\
NGC5121   &  GD   &    0.024 $\pm$ 0.007     &  3.457 $\pm$ 0.003 &	 PC1  &        F606W &	   6359 &   1	&  27.3 & 340\tablenotemark{d}  & \\
NGC5194   &  LW   &    0.160 $\pm$ 0.000\tablenotemark{c}  &  1.198 $\pm$ 0.006  &	  PC1  &       F547M &	   5123 &   4	&  22.1 & 125 & \\
NGC5236   &  C    &    0.190 $\pm$ 0.040\tablenotemark{b}  &  1.942 $\pm$ 0.027 &	 PC1  &        F814W &	   8234 &   5	&  7.7 &  251 & \\
NGC5248   &  TW   &    0.269 $\pm$ 0.064     &  1.717 $\pm$ 0.027 &	 PC1  &        F814W &	   6738 &   4	&  2.7 &  92\tablenotemark{d}  & \\
NGC5427   &  LW   &    0.231 $\pm$ 0.074     &  2.327 $\pm$ 0.010 &	 PC1  &        F606W &	   5479 &   5	&  22.7 & 376\tablenotemark{d}  &    LGD \\
NGC5643   &  GD   &    0.415 $\pm$ 0.013     &  1.150 $\pm$ 0.007 &	 PC1  &        F606W &	   8597 &   5	&  38.1 & 312  &    LGD \\
NGC6217   &  C    &    0.360 $\pm$ 0.010\tablenotemark{c}  &  6.472 $\pm$ 0.171  &	  PC1  &       F606W &	   5479 &   4	&  23.9 & 210 &    SB(rs) \\
NGC6221   &  C    &    0.436 $\pm$ 0.112     &  5.053 $\pm$ 0.010 &	 PC1  &        F606W &	   5479 &   5	&  19.4 & 200  &    SB(s) \\
NGC6300   &  CS   &    0.187 $\pm$ 0.002     &  2.505 $\pm$ 0.074 &	 PC1  &        F606W &	   5479 &   3	&  14.3 & 186  &    SB(rs), LGD \\
NGC6384   &  C    &    0.136 $\pm$ 0.020     &  0.910 $\pm$ 0.009 &	 PC1  &        F606W &	   6359 &   4	&  26.6 & 477  & \\
NGC6814   &  GD   &    0.070 $\pm$ 0.010\tablenotemark{a}  &  1.585 $\pm$ 0.049  &	  PC1  &       F606W &	   5479 &   4	&  22.8 & 205 & \\
NGC6951   &  LW   &    0.280 $\pm$ 0.040\tablenotemark{a}  &  2.036 $\pm$ 0.041  &	  PC1  &       F606W &	   8597 &   4	&  24.1 & 273 & LGD\\
NGC7098   &  C    &    0.200 $\pm$ 0.020\tablenotemark{b}  &  0.856 $\pm$ 0.003 &	 PC1  &        F555W &	   6633 &   1	&  29.1 & 345 & \\
NGC7213   &  TW   &    0.023 $\pm$ 0.002     &  1.124 $\pm$ 0.020 &	 PC1  &        F606W &	   5479 &   1	&  22 &   198  & \\
NGC7479   &  C    &    0.696 $\pm$ 0.060     &  2.022 $\pm$ 0.046 &	 PC1  &        F814W &	   6266 &   5	&  32.4 & 384  &    SB(s) \\
NGC7552   &  LW   &    0.395 $\pm$ 0.044     &  5.391 $\pm$ 0.267 &	 PC1  &        F606W &	   5479 &   2	&  19.5 & 192  &    SB(s) \\
NGC7727   &  LW   &    0.096 $\pm$ 0.024     &  1.222 $\pm$ 0.031 &	 PC1  &        F555W &	   7468 &   1	&  23.3 & 317  & \\
\enddata

\tablecomments{Literature data and measurements for the 75 galaxies in
  our sample.  The classification codes in column~2 are: GD: grand
  design nuclear spiral; TW: tightly wound spiral; LW: loosely wound
  spiral; CS: chaotic spiral; C: chaotic circumnuclear dust; N: no
  circumnuclear dust structure.  These codes are defined in
  \S\ref{sec:nuc}.  $Q_b$ (col.~[3]) is from \citet{laurikainen04}
  unless otherwise noted.  See \S\ref{sec:rms} for a discussion of the
  structure map rms $\sigma_{\mbox{\scriptsize sm}}$ (col.~[4]).  In
  column~5, ``HRC'' is the High Resolution Channel for ACS; the other
  detectors refer to different chips on WFPC2.  The Hubble T-types
  (col.~[8]) and SB classifications (col.~[11]) are from the RC3.  The
  distances (col.~[9]) are from \citet{tully88} and assume $H_0 =
  75$~km~s~$^{-1}$~Mpc$^{-1}$.  The solid white circles in
  Fig.~\ref{fig:stmaps} have radius $r_c$ (col.~[10]) as defined in
  eqn.~(\ref{eqn:rc}). LGD (col.~[11]) refers to ``large grand design''
  structure; see \S\ref{sec:gd} for details.  }

\tablenotetext{a}{\citet{block04}}
\tablenotetext{b}{\citet{buta01}}
\tablenotetext{c}{\citet{laurikainen02}}
\tablenotetext{d}{$r_c = r_{\mbox{\scriptsize eff}}(\mbox{bulge})$}
\end{deluxetable}

\end{center}
\clearpage
For each galaxy, we define a critical ``central'' radius $r_c$
by
\begin{equation}\label{eqn:rc}
r_c = \min\{ r_{\mbox{\scriptsize eff}}(\mbox{bulge}),\, 0.01\cdot
D_{25} \},
\end{equation}
where $r_{\mbox{\scriptsize eff}}(\mbox{bulge})$ is the bulge radius
reported in \citet{laurikainen04} and $D_{25}$ is the RC3 value of the
25th magnitude B-band isophotal diameter \citep{devaucouleurs91}.  If
\citeauthor{laurikainen04}\ give no bulge radius, or if the galaxy is
completely bulgeless, then $r_c = 0.01 D_{25}$.  As the bulge will
dominate the circumnuclear region, the minimum is used in the definition
of $r_c$ to guarantee that the central region is no larger than the
bulge.  Most (about 80\%) of the galaxies in the final sample have $r_c
= 0.01D_{25}$.  A circle of radius $r_c$, which corresponds to
30--550~pc projected, is included on the structure maps used for
classification (see \S\ref{sec:nuc}).

  Objects with high inclination (axis ratio $R_{25}$~$\leq$~$0.30$) or
low signal-to-noise ratio ($\lesssim$~10) are excluded from the sample.
We also discard low-resolution objects, given by
$r_c$~$\leq$~20~resolution elements, as well as ones with unfavorable
location on the chip or highly saturated centers.  In general, the
objects failing the resolution or signal-to-noise ratio cuts have no
clearly discernible coherent spiral arm structure.  All objects are
closer than 50~Mpc. Our final sample, summarized in
Table~\ref{tbl:data}, consists of the 75 galaxies which pass our
resolution and signal-to-noise ratio cuts; the T-type distribution of
the \citeauthor{eskridge02} sample is roughly maintained.

\section{Analysis}\label{sec:an}
We employ two methods of examining the nuclear dust content in galaxies.
The first is to look at what structures the dust forms, i.e., a
morphological classification, and the second is to simply study the
amount of dust structure in the central regions.  We describe in
\S\ref{sec:smap} how we use the structure map technique of
\citet{pogge02} to enhance dust structures on the scale of 1--15~pc.  We
then use these structure maps to classify the circumnuclear dust
morphology into six classes (four types of spirals, plus chaotic and no
structure), as described in \S\ref{sec:nuc}.  In \S\ref{sec:rms} we
describe how we estimate the relative amounts of central dust structure
in different galaxies.  Finally, in \S\ref{sec:monte}, we discuss how we
statistically compare pairs of distributions of galaxies.

\subsection{Structure Maps}\label{sec:smap}
As it is difficult to obtain a large sample of high quality, high
resolution multi-band (e.g., both $V$ and $H$) data, we used the
``structure map'' technique developed by \citet{pogge02} to enhance the
observable dust structure in each galaxy.  Structure maps can be thought
of as ``single-band color maps;'' while requiring data in only one band,
they provide a sharp contrast between dust and star-forming regions.
\citet{martini99b} note that the regions of highest contrast in $V-H$
color maps are primarily due to absorption by dust structures in the $V$
image alone, while the smoother $H$-band images trace the underlying
stellar population.  Therefore, the $H$-band images serve mostly to
subtract larger-scale surface brightness variations from the $V$-band
images in the $V-H$ color maps.  By making structure maps from images
taken at visible wavelengths, it is possible to enhance the structure due to
dust and emission regions.  Furthermore, as visible-wavelength images
from HST have higher resolution than those in the near infrared (both
being diffraction limited), a structure map can make the most of high
quality data.

 The structure map is mathematical defined as
\begin{equation}\label{eq:stmap}
S = \left[ \frac{I}{I \otimes P} \right] \otimes P^t ,
\end{equation}
where $S$ is the structure map, $I$ is the image, $P$ is the point
spread function (PSF), $P^t$ is the transform of the PSF, and $\otimes$
is the convolution operator.  Structure maps bring out variations on the
smallest resolvable scale of the image, i.e., that of the PSF.  For our
objects, this corresponds to a projected scale of 1--15pc.  The PSFs are
modelled using the TinyTim software \citep{krist99}.

\begin{figure}
\plotone{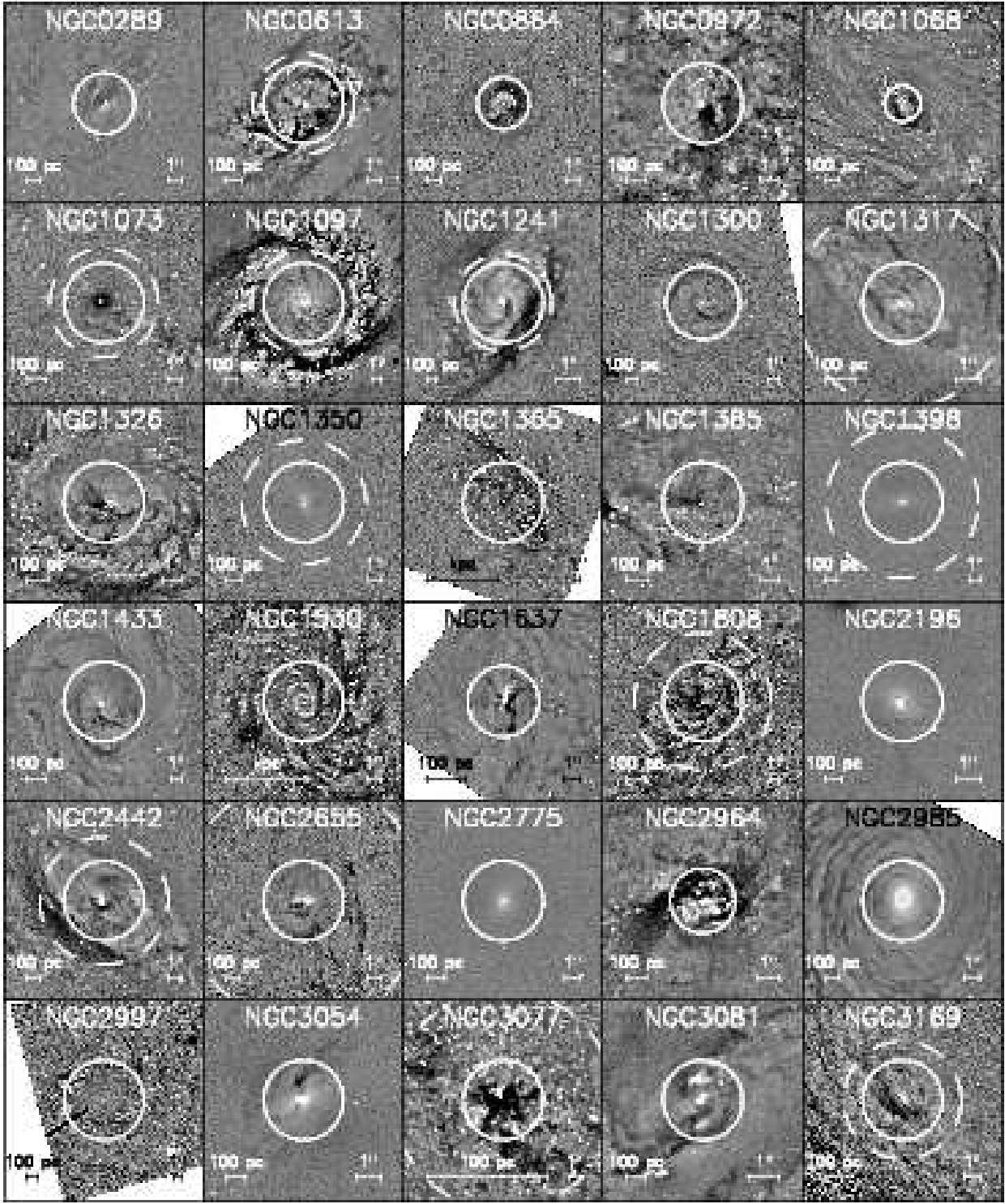}
\caption{Structure maps for 75 galaxies with measured barstrengths.
  Each panel shows the inner 5\% of $D_{25}$ from the RC3 catalog.  The
  solid white circles are given by a radius of $r_c$, as defined in
  eq.~(\ref{eqn:rc}).  In the case that $r_{\mbox{\scriptsize
  eff}}(\mbox{bulge}) > r_c$, the bulge radius is marked by a dashed
  white circle; these dashed circles were not included on the structure
  maps used for classification, but are merely given here for reference.
  Dark regions are due to dust, while bright regions are due to
  emission.  Several images (e.g., NGC2997) also include the ACS
  coronographic finger.  North is up and East is to the left.
\label{fig:stmaps}}
\end{figure}

Formally, structure maps are similar to the second-order iteration of
the Richardson-Lucy (R-L) image correction \citep{richardson72, lucy74}.
At this step in the image restoration, the first-order, smooth structure
of the image has been removed, but the structures on the scale of the
size of the PSF remain.  Structure maps for the 75 galaxies in our
sample are given in Figure~\ref{fig:stmaps}; in all of the structure
maps presented in this paper, the dusty regions appear dark while
enhanced stellar light and emission-line regions are bright (e.g., the
F606W filter admits several bright emission lines.)

\clearpage
\clearpage
{\plotone{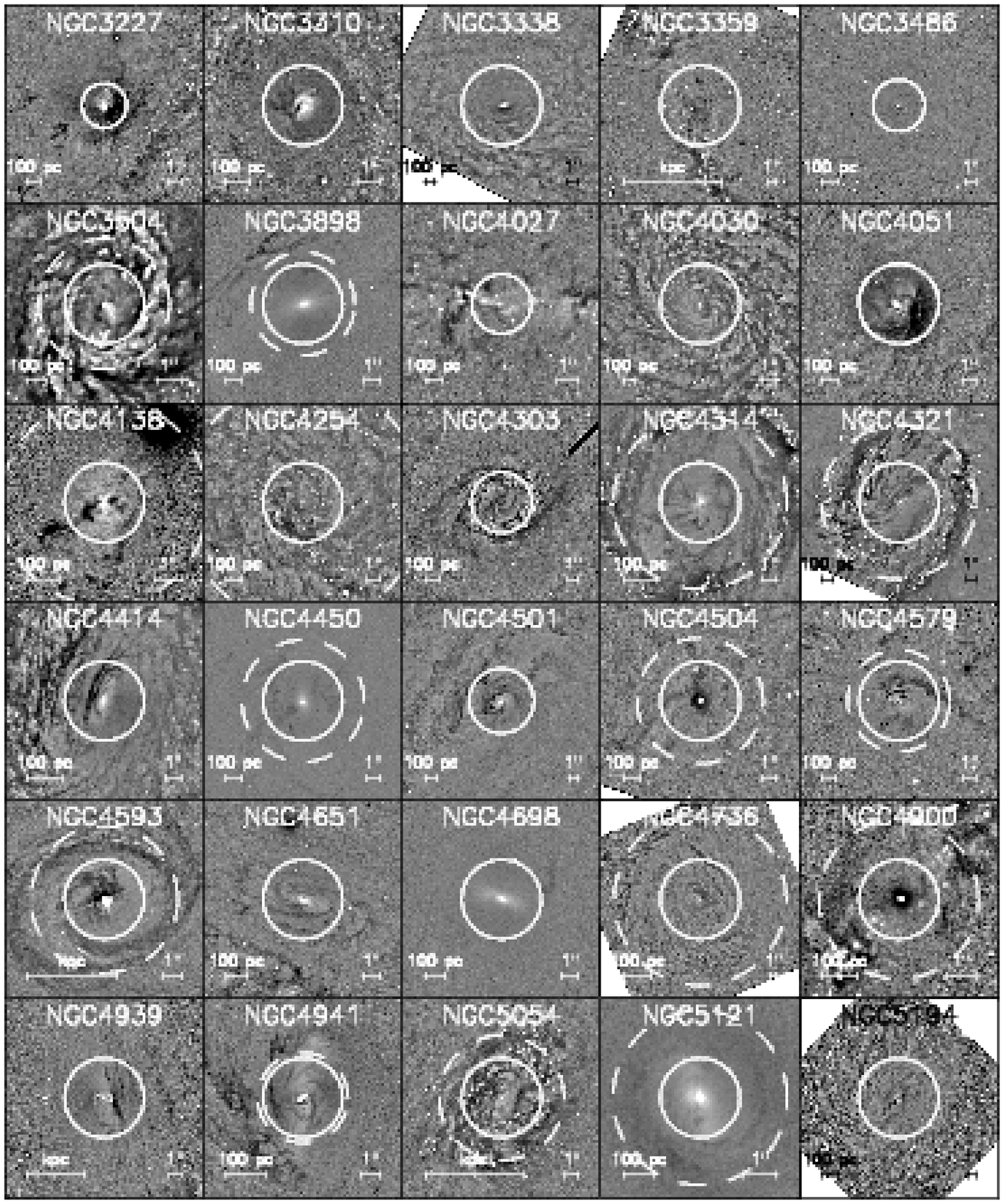}}\\[5mm]
\centerline{Fig. 1. --- Continued.}
\clearpage
{\plotone{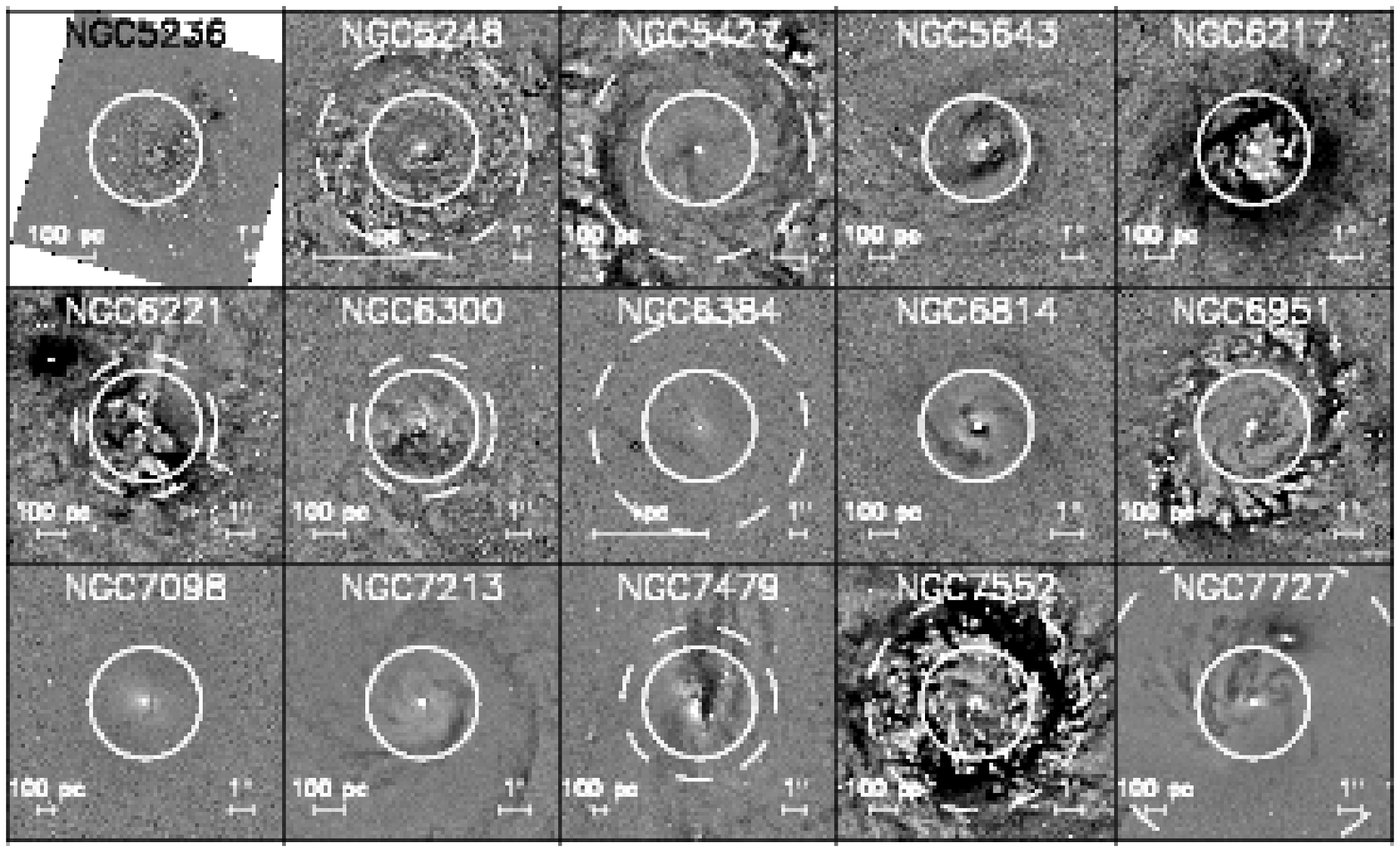}}\\[5mm]
\centerline{Fig. 1. --- Continued.}
\clearpage

\subsection{Nuclear Dust Classification}\label{sec:nuc}
We classify the galaxies according to a refined version of the nuclear
classification scheme presented in \citet{martini03a}.  The main
distinction between the scheme used here and the original one is that we
have a stronger emphasis on the central-most regions of the galaxy;
other differences between the scheme used here and the one in
\citet{martini03a} are discussed below.  The structure maps used for
classification are scaled to have a width and height equal to 5\% of
$D_{25}$.  The classification scheme uses six categories:
\begin{description}

\item[Grand Design (GD):] Two spiral arms symmetric about a 180\dg\
  rotation, at least one of which extends all the way to the unresolved
  nucleus of the galaxy, and at least one of which is a dominant feature

\item[Tightly Wound (TW):] Coherent spiral arm structure over a large
  range in radius; pitch angle less than or equal to 10\dg

\item[Loosely Wound (LW):] Coherent spiral arm structure over a large
  range in radius; pitch angle greater than 10\dg

\item[Chaotic Spiral (CS):] Unambiguous evidence for spiral arm
  structure with a unique sense of chirality, but not coherent over a
  large range in radius

\item[Chaotic (C):] Dust structure with no evidence of spiral structure

\item[No Structure (N):] No discernible dust structure within the
  central region

\end{description}

\begin{figure}
\plotone{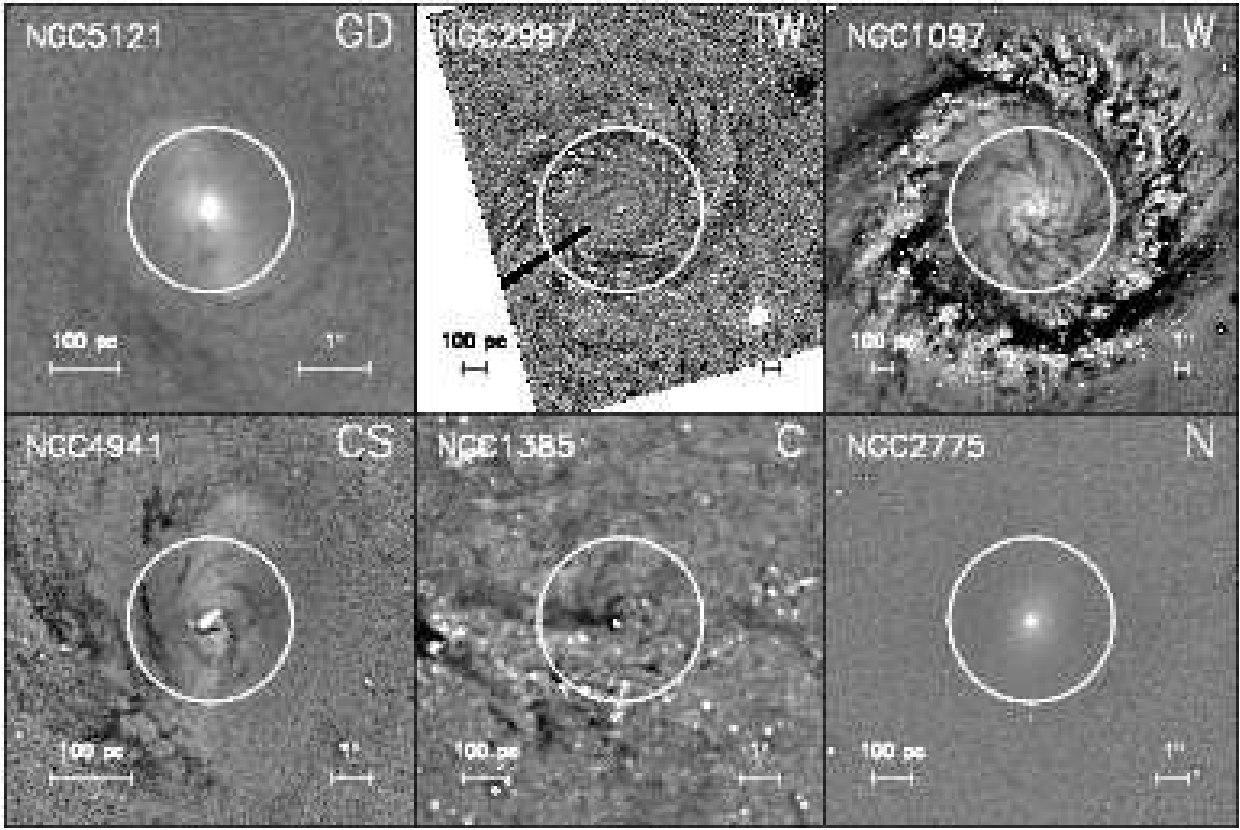}
\caption{Structure maps of prototypes for the six circumnuclear
  morphology classes discussed in \S\ref{sec:nuc}.  The white circles
  are given by a radius of $r_c$, as defined in eq.~(\ref{eqn:rc}).
  Dark regions are dust, while bright regions are emission.
\label{fig:proto}}
\end{figure}

Examples of each classification are given in Figure~\ref{fig:proto}, and
the classifications for all of the galaxies in our sample are given in
Table~\ref{tbl:data}.  There are several subtle, yet important,
refinements to the original classification scheme.  First, the
classification is based on a region that scales with the size of the
galaxy (the central 5\% of $D_{25}$) rather than within the central
$19.6''$ used by \citet{martini03a}.  Most changes in classification can
be attributed to this more physically defined focus.  The largest
difference in an individual class is the requirement that for an object
to be classified as GD, at least one of the arms must be observed to
extend to the nucleus of the galaxy.  For this reason, some galaxies
previously classified as GD now have a different classification (e.g.,
NGC1365).  We comment on this distinction in greater detail in
\S\ref{sec:gd}.
\clearpage
\clearpage
We also quantify the distinction between tightly and loosely wound
circumnuclear spirals by setting 10\dg\ as the delineation between
``small'' and ``large'' pitch angles. This division roughly corresponds
to that of Sb galaxies (Kennicutt 1981).  Previously, the
differentiation between TW and LW was done purely by eye, and also took
into account arm morphology.  Specifically, for a nuclear spiral to be
classified as ``tightly wound,'' the arms had to be traceable through at
least one complete revolution, and therefore the classification was
potentially subject to signal-to-noise ratio bias (for an arm to remain
coherent about an entire revolution, and thus be classified as TW, a
higher SNR was needed).  We measure pitch angles with an interactive
PGPlot program which fits logarithmic spirals to a deprojected image of
the structure map.  Five points of increasing radius on a single
prominent spiral arm of the galaxy are chosen within an annulus given by
25--75\% of the bulge radius, or of 1\% of $D_{25}$ if the galaxy is
either bulgeless or there is no available effective bulge radius
data.\footnote{This radius is distinct from $r_c$ in that if
\citet{laurikainen04} report a non-zero bulge radius, then it is used,
rather than taking the minimum of $r_{\mbox{\scriptsize
eff}}(\mbox{bulge})$ and $0.01D_{25}$.  This is because at larger radii,
there is more likely to be a coherent arm with a measurable pitch
angle.}  The program then fits logarithmic spirals to each consecutive
pair of points, averaging the resulting four pitch angles to give the
pitch angle for the entire arm.  As found by \citet{kennicutt81} for
large-scale spirals, the nuclear spirals are not well fit by logarithmic
spirals---a given arm does not have constant pitch angle as a function
of radius.  Still, this method yields an ``average'' pitch angle for the
arm.  The observed non-constant pitch angle along a spiral arm implies
that the region is experiencing differential rotation which is
inconsistent with constant rotational velocity \citep{maciejewski04a}.  We
also note that the pitch angle often varies between different arms in a
single galaxy.  Notwithstanding these caveats, we find that this method
suffices for determining whether or not the pitch angle is above or
below 10\dg, thus differentiating between TW and LW nuclear spirals.

Finally, the ``central region'' used in classifying the N class of
objects was previously defined as a ``few hundred parsecs''
\citep{martini03a}.  We have clarified this so that the central region
is specific to each galaxy, namely, a circle of radius $r_c$.  While in
general the entire region within $0.05D_{25}$ is used for
classification, the region within $r_c$ is also used as a guide in
deciding between different classifications; if a galaxy is seen to have
a different morphology at small radii than at large radii, such as is
seen in NGC1068, the classification at small radii is used.  This
decision also led to changes in the classifications of several galaxies
from those reported in previous papers. Though this distinction is
subject to the definition of $r_c$, only two galaxies, NGC1068 and
NGC3227, would have their classifications changed (LW to C and CS to C,
respectively) if $r_c$ were defined to be 1\% of $D_{25}$ for all
galaxies, rather than the minimum used in Equation~(\ref{eqn:rc}).
Despite this, we choose to keep the more physically motivated definition
of $r_c$ for reasons discussed in \S\ref{sec:data}.

\subsection{Dust Contrast}\label{sec:rms}
To quantify the amount of relative structure galaxies, we calculated the
root mean squared (rms) of the pixel-to-pixel variations of the
galaxies' structure maps within $r_c$. The more dust structure or star
formation in a galaxy, the higher the dust contrast in the structure
map, and therefore the higher the rms.  We assign an uncertainty to the
rms by calculating the standard deviation of 11 measurements of the rms
within a circular aperture ranging in radius from $0.95r_c$ to $1.05r_c$
in increments on $0.01r_c$.  We further define
$\sigma_{\mbox{\scriptsize sm}}$ to be the average of these 11 measured
rms values.  The variation on $\sigma_{\mbox{\scriptsize sm}}$ is found
to be no more than 5\% for any object.  The measured rms values and
uncertainties for each galaxy are given in Table~\ref{tbl:data}.

\begin{figure}
\plotone{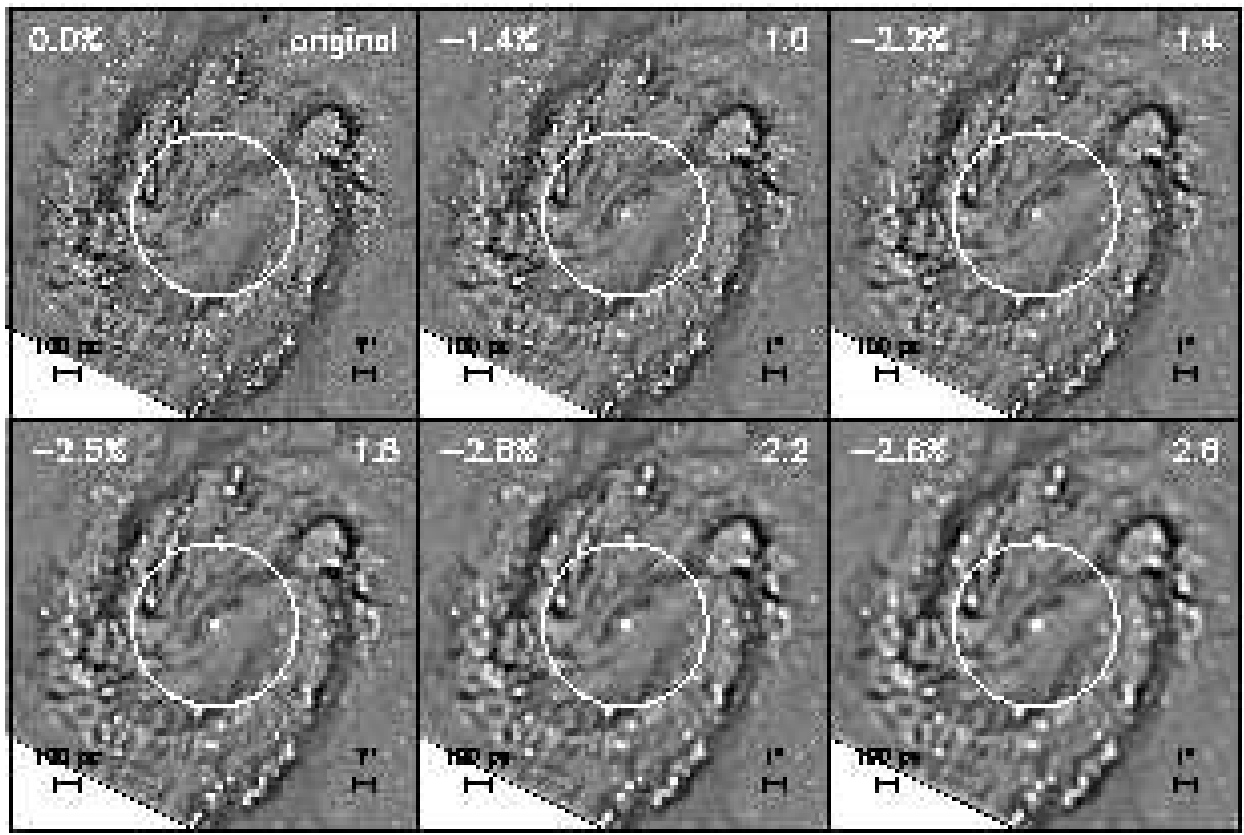}
\caption{Degraded structure maps of NGC4321 for testing the dependence
  of structure map rms $\sigma_{\mbox{\scriptsize sm}}$ on resolution.
  The percents in the upper left-hand corners are the percent change in
  $\sigma_{\mbox{\scriptsize sm}}$ relative to the original structure
  map.  The numbers in the upper right-hand corners are the standard
  deviations in pixels of the Gaussians with which both the image and
  the PSF were convolved before construction of the structure map.  The
  white circles have a radius of $r_c$, as defined in
  eq.~(\ref{eqn:rc}).  The resolution of the original image is $0.06''$,
  or 4.9~pc, while the resolution of the 2.6-$\sigma$ Gaussian convolved
  image is $0.13''$, or 10.8~pc.  This is equivalent to moving the
  galaxy from 16.8~Mpc to 37~Mpc.  Relative to the measured uncertainty
  on the original $\sigma_{\mbox{\scriptsize sm}}$ (2.1\%), there is no
  significant change in $\sigma_{\mbox{\scriptsize sm}}$ with
  resolution.
\label{fig:rmsdegrade}}
\end{figure}

Because the structure map is sensitive to structures on the scale of the
PSF, rather than some physical scale, there may be a strong
distance-dependent bias in $\sigma_{\mbox{\scriptsize sm}}$.  We
therefore tested how $\sigma_{\mbox{\scriptsize sm}}$ changes for a
given galaxy upon changing the resolution using a test sample of nine
galaxies with high enough resolutions to meet our resolution and
signal-to-noise cuts (see \S\ref{sec:data}) even after being convolved
with a Gaussian with a standard deviation of 2.6 pixels, which
corresponds to more than doubling their distance.  For each object, we
convolved both the image and the PSF with a Gaussian with a standard
deviation of 1.0--2.6 pixels (in 0.1 pixel increments), from which a set
of 17 degraded structure maps were constructed.  An example of this
degradation for NGC4321 is shown in Figure~\ref{fig:rmsdegrade}.  We
then calculated $\sigma_{\mbox{\scriptsize sm}}$ (and uncertainty, as
described above) for each of these degraded structure maps.  While
$\sigma_{\mbox{\scriptsize sm}}$ did change for each object, the change
was found to be comparable with the calculated uncertainties for any
given structure map.  This implies that at these scales, the dust
structure is roughly scale-invariant, in agreement with the findings of
\citet{elmegreen02} that the power spectra of dust structure becomes
relatively shallow at small scales.  We thus conclude that
$\sigma_{\mbox{\scriptsize sm}}$ does not contain a distance-dependent
bias and employ it below to compare different subsamples of galaxies.
\clearpage
\clearpage

\subsection{Comparing $Q_b$ and $\sigma_{\mbox{\scriptsize sm}}$ Distributions}\label{sec:monte}
Once the galaxies are classified, we have sets of barstrengths $Q_b$ and
structure map rms values $\sigma_{\mbox{\scriptsize sm}}$ for each
nuclear classification.  We then want to answer the question of how
likely it is that the objects of a given classification are drawn from
the same parent distribution as other galaxies.  The standard method for
comparing two arbitrary distributions is the Kolmogorov-Smirnov (K-S)
test, which calculates the maximum vertical separation $D_{KS}$ between
two cumulative distributions.  The larger the K-S distance $D_{KS}$ for
two samples, the less likely the two distributions represent the same
parent population.  $D_{KS}$ only takes on discrete values, namely,
multiples of $1/(N_A+N_B)$, where $N_A$ and $N_B$ are the number of
elements in the two samples A and B under comparison.  To compare two
samples, we first calculate the distance $D_{KS}$ between the two
distributions.  Then, for 10,000 trials, we compare this $D_{KS}$ to the
distance between two samples of size $N_A$ and $N_B$ drawn from a
uniform distribution, yielding a probability that the two samples are in
fact drawn from the same distribution.  These probabilities do not
change significantly when errors on $Q_b$ or $\sigma_{\mbox{\scriptsize
sm}}$ are taken into account.  In the event that the two samples
intersect, we exclude the shared galaxies from both samples in order to
do the comparison.

In the few cases where both sample sizes small, we use the Wilcoxon
rank-sum test instead of the K-S test.  The Wilcoxon test compares two
samples A and B of size $N_A$ and $N_B$, with $N_A \le N_B$, where the
``true'' mean values of A and B are $\mu_A$ and $\mu_B$, respectively.
Letting the sum of the ranks in sample A be $w$, the probability that
$\mu_A > \mu_B$ is equal to the probability that the sum of the ranks of
$N_A$ randomly chosen elements (from a set of size $N_A + N_B$) is $\le
w$, while the probability that $\mu_A < \mu_B$ is equal to the
probability that the sum of ranks is $\ge N_A (N_A + N_B + 1) - w$.

\section{Results and Discussion}\label{sec:results}
\begin{figure}
\plotone{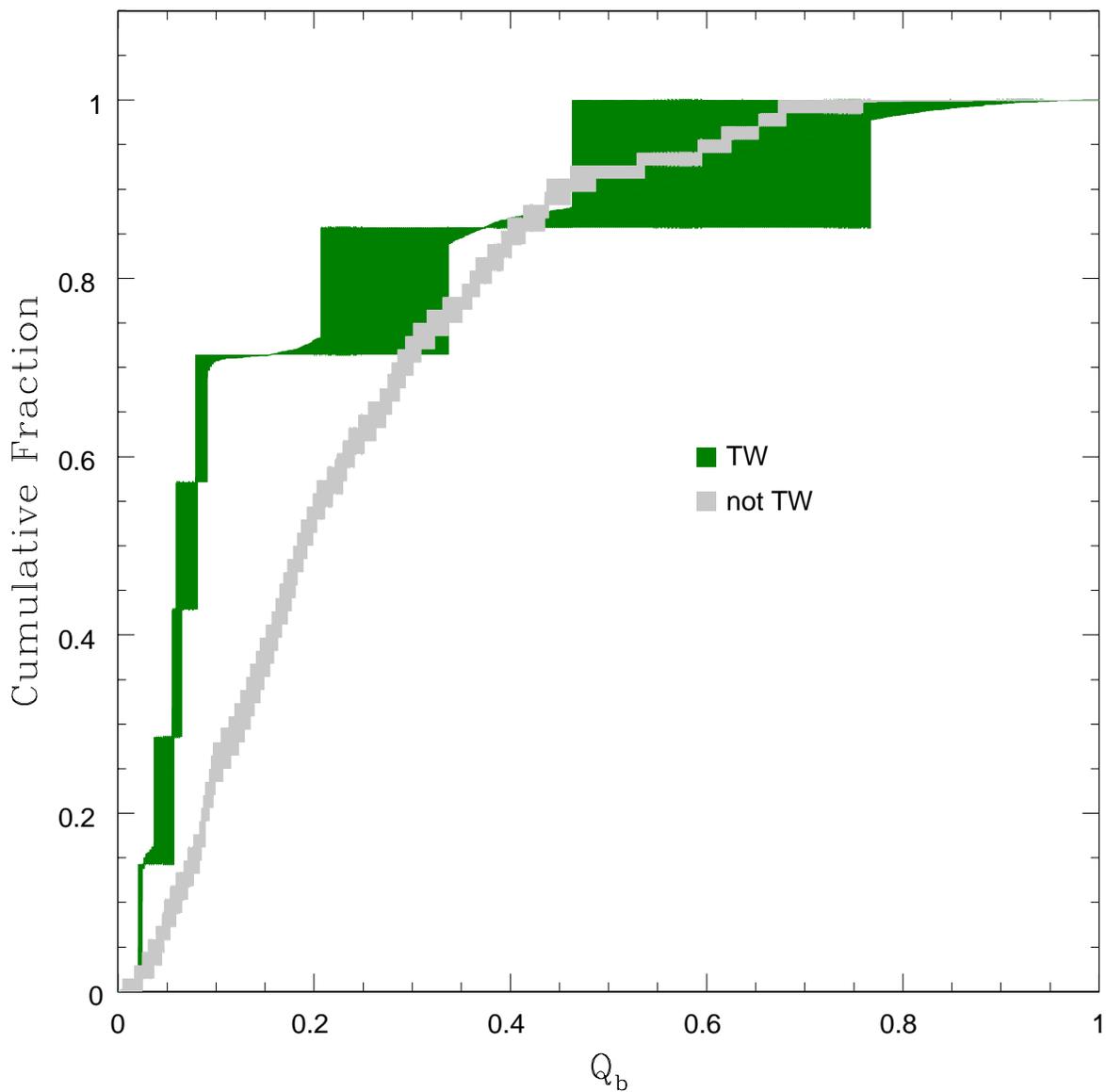}
\caption{Comparison of $Q_b$ for TW galaxies and all other galaxies.
  The vertical widths correspond to the central 68-percentile spread of
  the cumulative fraction and are due to the uncertainty in $Q_b$.  Galaxies with TW circumnuclear dust spirals
  are more weakly barred than typical galaxies, with a 2\% probability
  of being drawn from the same parent distribution.
\label{fig:tw}}
\end{figure}

\begin{figure}
\plottwo{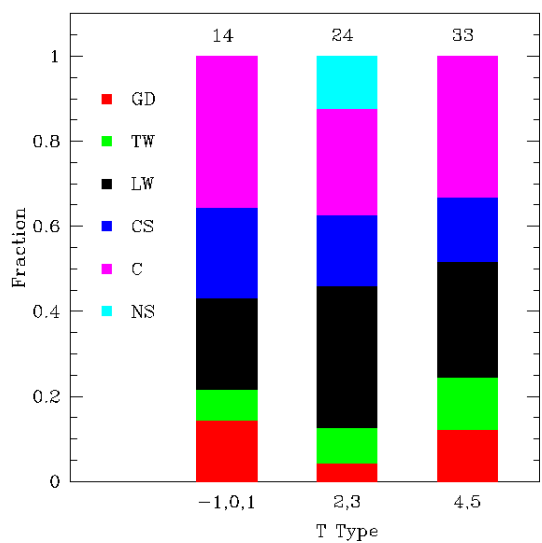}{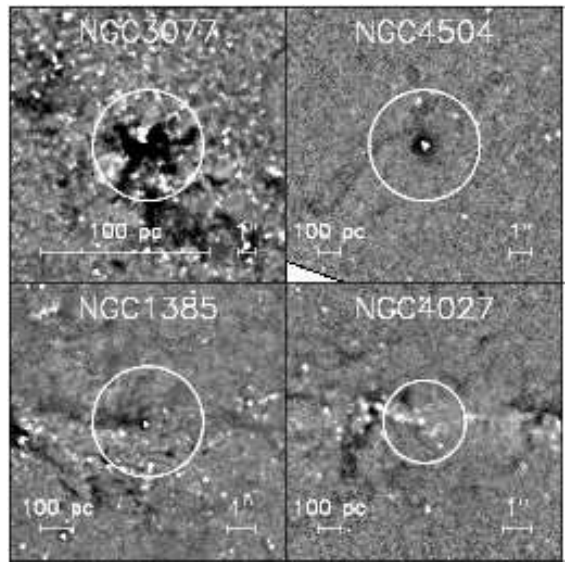}
\caption{{\em Left:} No connection is seen between Hubble T-type for
  T~$\le$~5 (Sc) and nuclear classification.  {\em Right:} Structure maps for
  the four galaxies in the sample with T~$\ge$~6.  The large scale
  morphologies from the RC3 catalog are: NGC3077, I0; NGC4504, Scd;
  NGC1385, Scd; NGC4027, Sdm.  All four of these galaxies are classified
  as having chaotic circumnuclear dust structure.
\label{fig:latet}}
\end{figure}

\begin{figure}
\plottwo{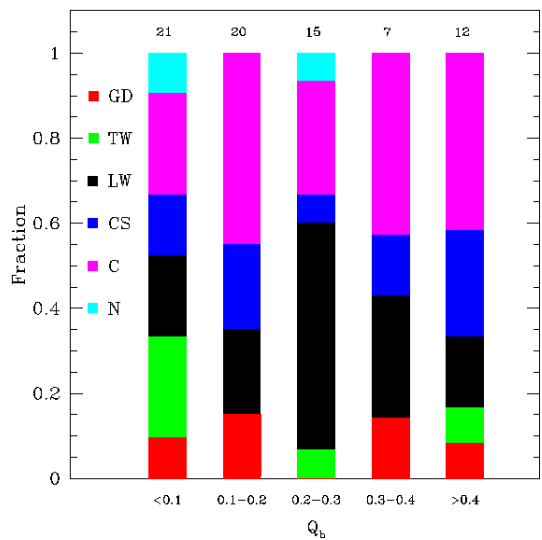}{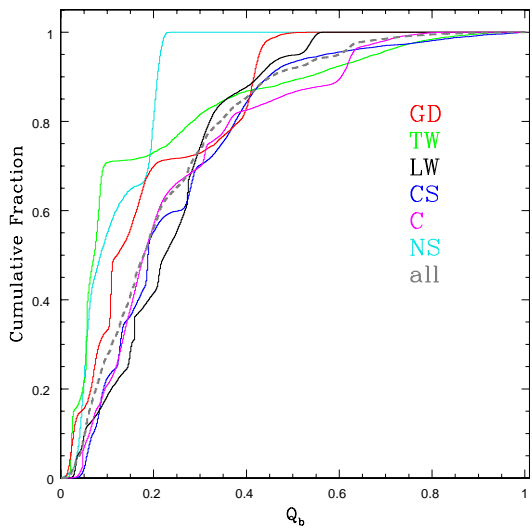}
\caption{Nuclear classifications and barstrength $Q_b$.  {\em Left:}
  Distribution of nuclear classification into different barstrength
  classes.  {\em Right:} Monte Carlo average cumulative distribution of
  barstrength in each nuclear class and total sample.
\label{fig:allqb}}
\end{figure}

Our classifications and measurements of $\sigma_{\mbox{\scriptsize sm}}$
are summarized in Table~\ref{tbl:data}.  There is a 2\% probability that
the tightly wound (TW) nuclear dust spirals have the same underlying
distribution of barstrength as the rest of the sample.  As shown in
Figure \ref{fig:tw}, TW nuclear spirals are found primarily---although
not exclusively---in weakly barred galaxies.  As the TW class is defined
to be those dust spirals with pitch angle $\le$~10\dg, this is
consistent with the idea that a galaxy which is axisymmetric at large
scales (i.e., has a small $Q_b$) will either have high differential
rotation or simply that the central regions of these galaxies cannot be
highly disrupted \citep{maciejewski04b}.  All four late-type galaxies in
our sample (T~$\ge$~6) have chaotic circumnuclear dust structure, as
shown in the right-hand panel of Figure~\ref{fig:latet}.  No such
differences were found for any other nuclear class,\footnote{While all
three N class galaxies have a T~$=2$ (corresponding to a Hubble type of
Sab), this is not significant.} as shown in the left-hand panel of
Figure~\ref{fig:latet} and Figure~\ref{fig:allqb}.  As previous studies
have suggested a connection between large-scale bars and grand design
nuclear spirals, we begin our discussion of strongly barred galaxies
below with a consideration of grand design nuclear spirals.
\clearpage

\clearpage

\subsection{Circumnuclear Grand Design Spirals}\label{sec:gd}
\begin{figure}
\plotone{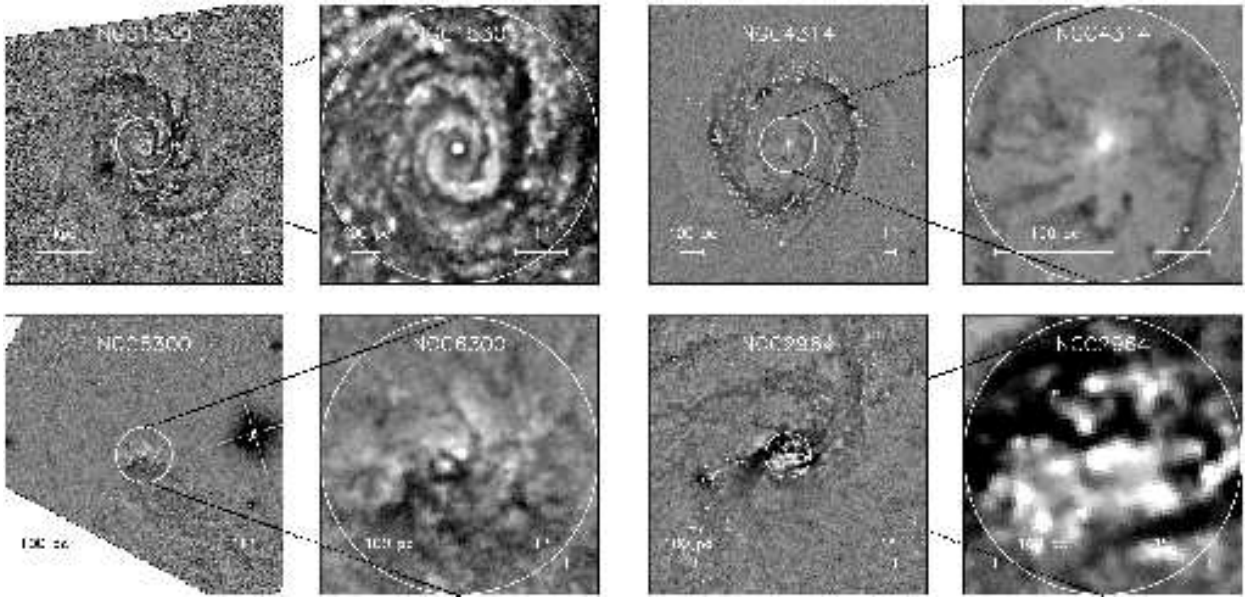}
\caption{Four examples of galaxies with LGD (large grand design) but not
  SGD (small grand design) structure.  {\em Left:} Width and height are
  given by $0.1D_{25}$.  {\em Right:} Width and height are given by
  $r_c$.  The classifications are: NGC1530, TW; NGC4314, LW; NGC6300,
  CS; NGC2964, C.  The white circles in both panels have radius $r_c$.
  Dark regions are due to dust, while bright regions are due to
  emission.  North is up and East is to the left.
\label{fig:gd}}
\end{figure}

Hydrodynamic simulations suggest that large-scale bars can lead to the
formation of grand design nuclear spirals \citep{patsis00, englmaier00},
and several observational studies support this hypothesis
\citep{pogge02, martini03b}.  With our focus on the very central regions
of nuclear spirals, we find that this is in fact not the case: the grand
design nuclear spirals described in \S\ref{sec:nuc} are {\em not}
preferentially found in strongly barred galaxies.  We find, however,
that symmetric two-arm spiral structure is common at larger scales in
strongly barred galaxies.  To differentiate between circumnuclear grand
design spirals found at different scales, we hereafter refer to grand
design structure at small radii, defined as GD in \S\ref{sec:nuc}, as
SGD structure (small grand design), and prominent grand design structure
at larger radii (within 10\% of $D_{25}$, as shown in
Figure~\ref{fig:gd}) as LGD structure (large grand design).

\begin{figure}
\plottwo{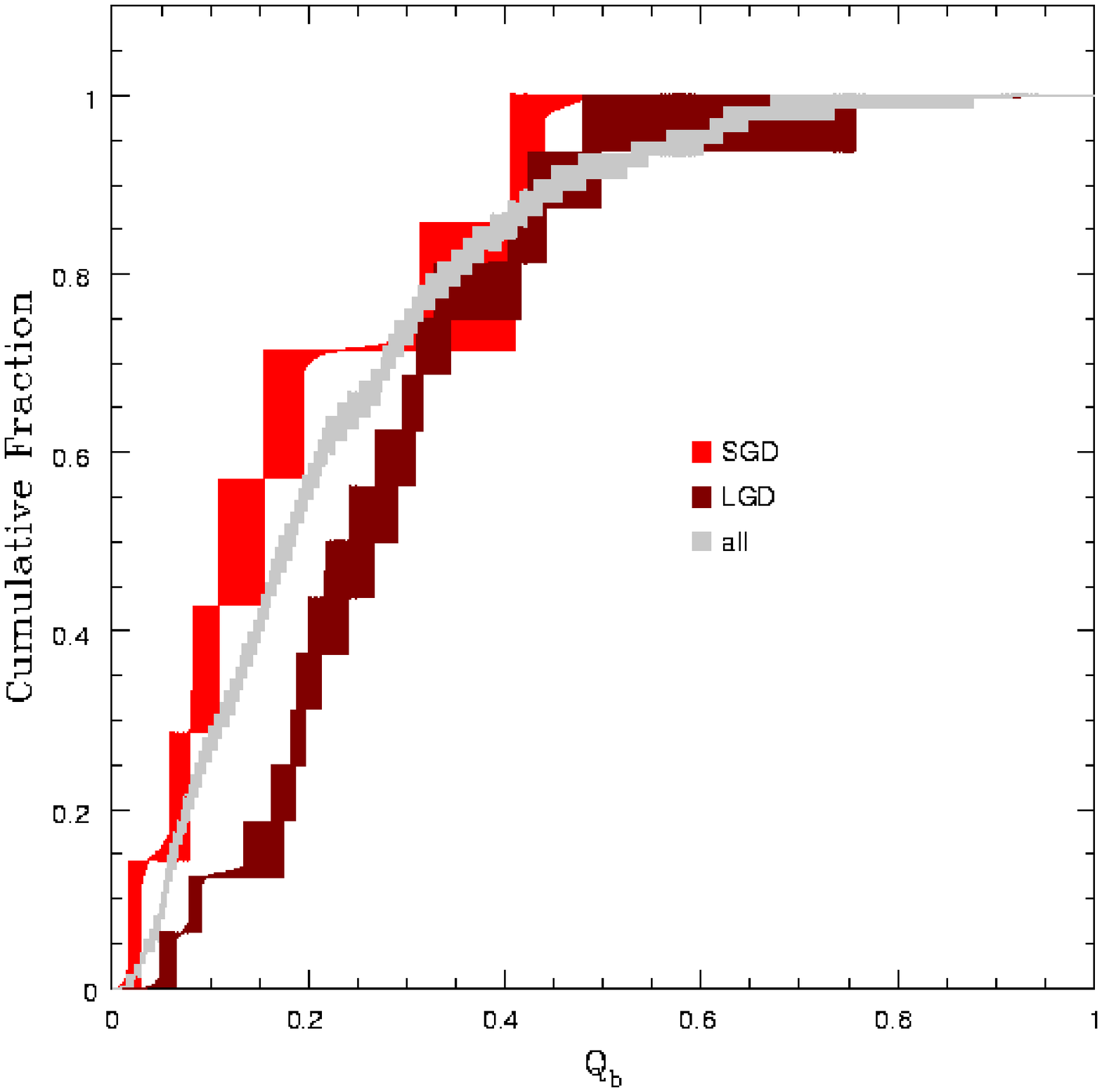}{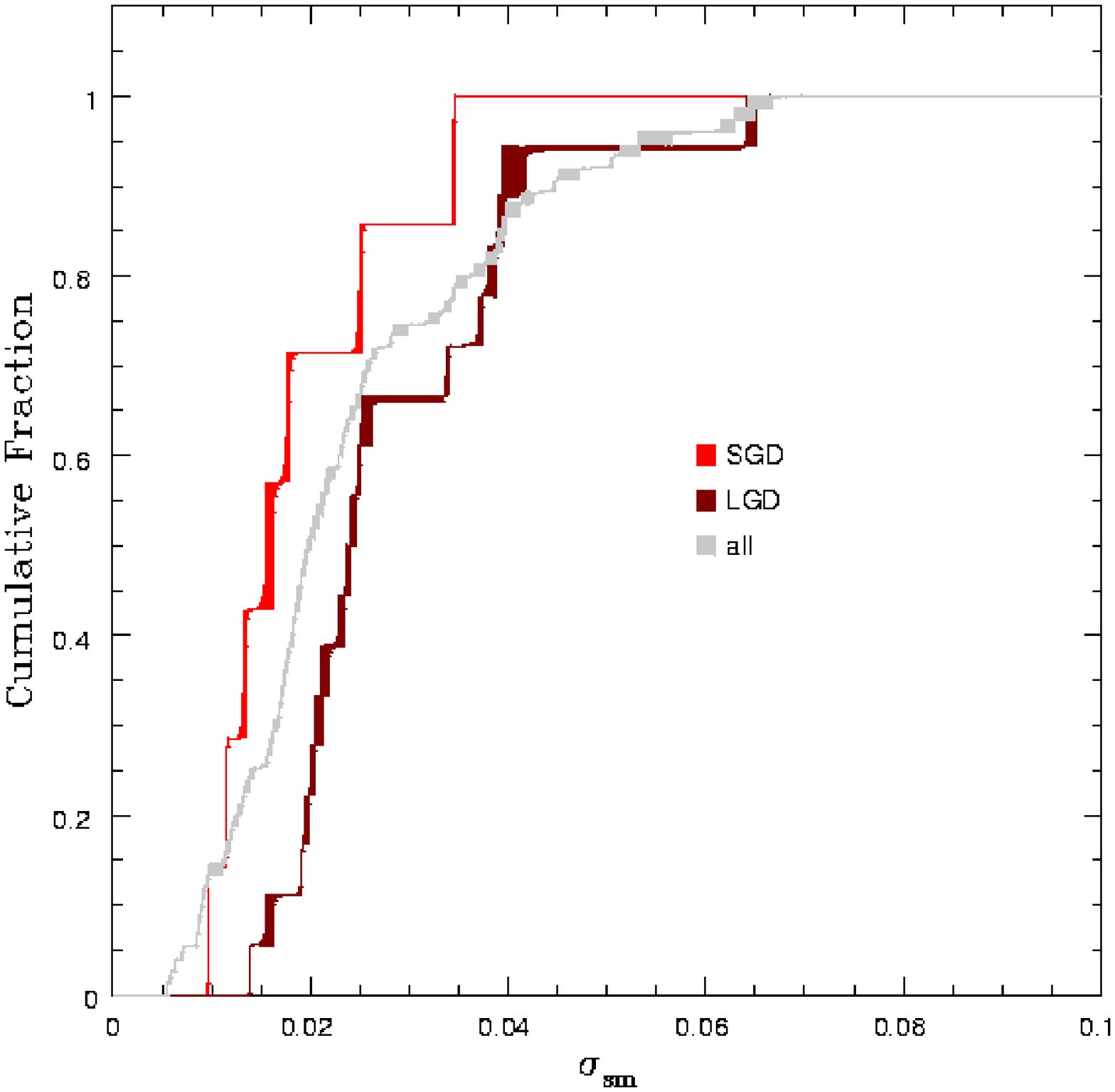}
\caption{Comparisons of barstrengths $Q_b$ and structure map rms
$\sigma_{\mbox{\scriptsize sm}}$ for SGD and LGD galaxies.  The vertical
widths correspond to the central 68-percentile spread of the cumulative
fraction and are due to the uncertainty in $Q_b$ and
$\sigma_{\mbox{\scriptsize sm}}$.  SGD galaxies are less
strongly barred (left) and have less dust structure (right) than LGD
galaxies.  There is a 0.4\% probability that the $Q_b$ are
drawn from the same parent population, and 3\% probability that the rms
values are drawn from the same parent distribution.
\label{fig:lgdvsgd}}
\end{figure}

Figure~\ref{fig:gd} shows four examples of galaxies for which grand
design structure is evident at larger scales, but is not present at
smaller scales.  Seven galaxies (9\% of the entire sample) have SGD
structure, while twenty (26\%) have LGD structure.  This larger-scale GD
structure extends to the center of the galaxy in only two (10\%) of
these twenty galaxies.  As can be seen in Figure~\ref{fig:lgdvsgd}, LGD
galaxies are preferentially more strongly barred than SGD galaxies.  A
K-S test excluding the two shared galaxies reveals a 3\% probability
that SGD and LGD barstrengths are drawn from the same parent
distribution.  We further find a 0.4\% probability that LGD galaxies
have the same underlying distribution of $Q_b$ as the other 59 galaxies
in the sample; galaxies with LGD nuclear spirals are more strongly
barred than other galaxies.  It is interesting to note that the two LGD
galaxies with the smallest $Q_b$ (NGC4941 and NGC1317) also have an
obvious nuclear bar within the grand design structure
(\citealp{erwin04} and \citealp{greusard00}, respectively).
\clearpage

\clearpage
Visually, SGD galaxies seem to have less circumnuclear dust than their
LGD counterparts.  A K-S test comparing the $\sigma_{\mbox{\scriptsize
sm}}$ values of the two samples supports this impression, yielding a 3\%
probability that the two are drawn from the same parent distribution
(see Figure~\ref{fig:lgdvsgd}).  This is in agreement with the overall
trend discussed below in \S\ref{sec:rings} that more strongly barred
galaxies have dustier circumnuclear regions, suggesting that the gas and
dust funnelled to the center of a barred galaxy may eventually suffice
to disintegrate a symmetric two-armed spiral in the central few hundred
parsecs.  The most likely mechanism for erasing the small-scale spiral
structure is self-gravity, and later, energy feedback from substantial
star formation \citep[e.g.,]{wada02}.  We note that many of the galaxies
with the largest $\sigma_{\mbox{\scriptsize sm}}$ (e.g., NGC1808,
NGC2964, NGC6217 and NGC7552) are well-known starbursts or nuclear
starbursts.  This indicates that there may be sufficient feedback from
star formation to shape the structure of the circumnuclear ISM.

\subsection{Circumnuclear Rings and Strongly Barred Galaxies}\label{sec:rings}
The material funnelled inward by the large-scale bar may not reach
the central region at all, but instead form a circumnuclear starburst ring.
If bars prompt rings of circumnuclear star formation,
then star-forming rings should be more prevalent in strongly
barred galaxies than in weakly barred ones.  Observations do suggest that
barred galaxies have a higher incidence of nuclear starburst rings than
unbarred galaxies \citep[see][and references therein]{kormendy04}.  We
offer here more evidence that this is in fact the case.

Of the 75 galaxies studied, eight have circumnuclear starburst rings
visible at radii less than or equal to 5\% of $D_{25}$.  All eight
exhibit LGD structure,\footnote{The LGD structure in NGC1300 and
NGC0864, while visible on the structure maps, is more prominent on the
$V-H$ color maps of \citet{martini03a}.  This is because in these two
cases, the relevant dust lanes are larger than the PSF, and thus have
low contrast on the structure map.} as shown in
Figure~\ref{fig:rings_zoom}.  One interesting transition case is
NGC5427, where the LGD spiral arms appear to become the incomplete
starburst ring itself.  NGC4314 also exhibits this morphology, although
the ring is more complete.  The LGD dust lanes enter a number of other
rings (e.g., NGC1097 and NGC6951), but these ``breaks'' may simply be
due to dust attenuation of a complete ring.  As seen in
Figure~\ref{fig:stmaps}, the effective bulge radius and the
circumnuclear ring radius for each of these galaxies apparently
coincide.  Of these eight galaxies, six have loosely wound circumnuclear
spirals within the ring, while the other two (NGC864 and NGC1326) have
chaotic circumnuclear spirals.  Only 32 of the 75 galaxies are LW or CS;
according to a binomial distribution, the probability of randomly
selecting 8 galaxies that are restricted to these two classes is
0.062\%.  The circumnuclear morphology within a circumnuclear ring is
therefore distinct from the ring itself.  Visually, the rings appear
highly turbulent and almost flocculent, while the dust structure in
their interiors is much smoother.

\begin{figure}
\plotone{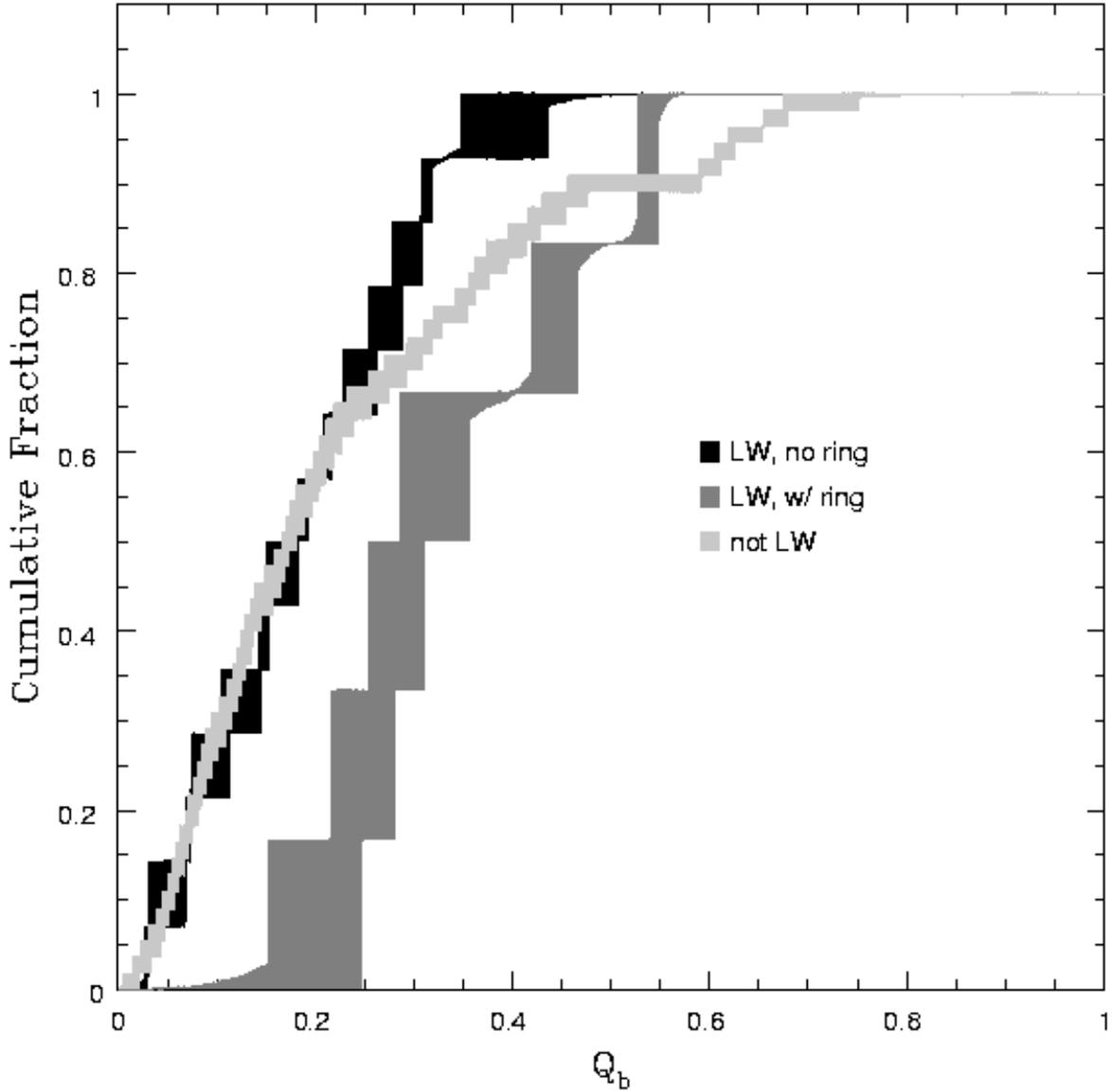}
\caption{Comparison of $Q_b$ for LW galaxies with and without
    circumnuclear rings.  The vertical widths correspond to the central
    68-percentile spread of the cumulative fraction and are due to the
    uncertainty in $Q_b$.  Galaxies with circumnuclear rings are found
    to be more strongly barred than typical galaxies, with 1.5\%
    probability of being drawn from the same parent population.
\label{fig:ringsQ}}
\end{figure}

\begin{figure}
\plotone{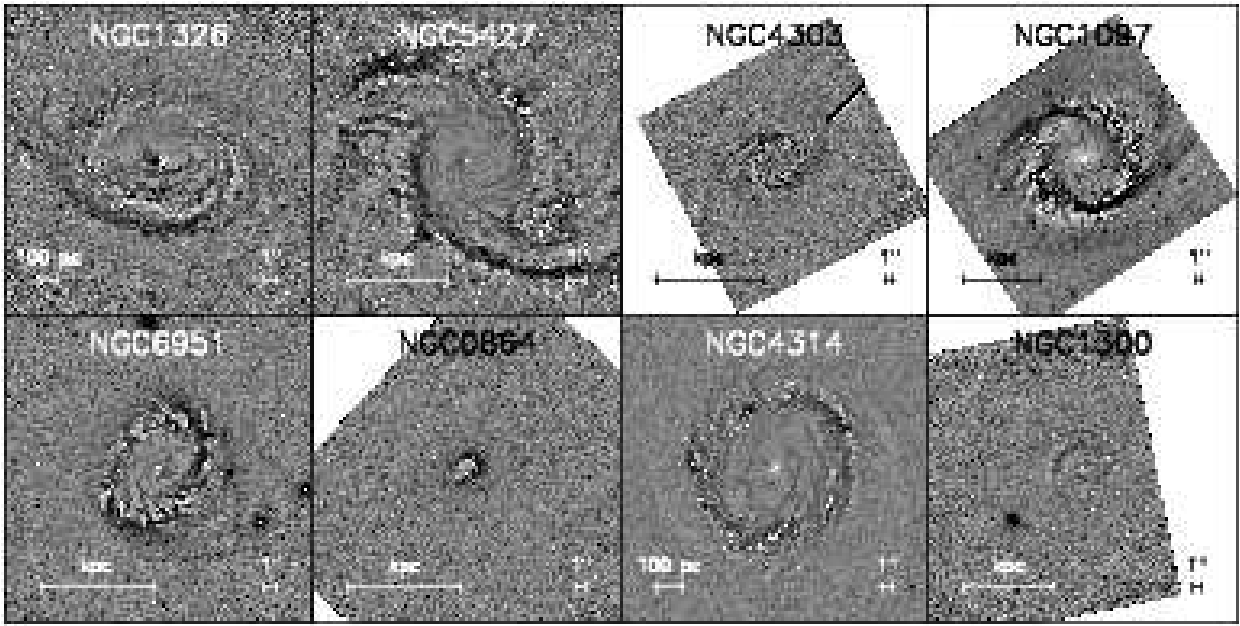}
\caption{Structure maps of the eight galaxies in our sample with
  circumnuclear rings in order of increasing $Q_b$.  All eight have LGD
  structure.  Width and height are given by $0.1D_{25}$.
\label{fig:rings_zoom}}
\end{figure}

The galaxies with circumnuclear rings are also, as a group, more
strongly barred than typical galaxies. The K-S test gives a probability
of 1.5\% that the ringed galaxies have the same barstrength distribution
as the other 67 galaxies, and a probability of 3\% that the LW ringed
galaxies have the same distribution of barstrengths as the other
fourteen LW galaxies.  The barstrengths of the entire, LW, and ringed
galaxy samples are shown in Figure~\ref{fig:ringsQ}.
\clearpage
\clearpage

\begin{figure}
\plotone{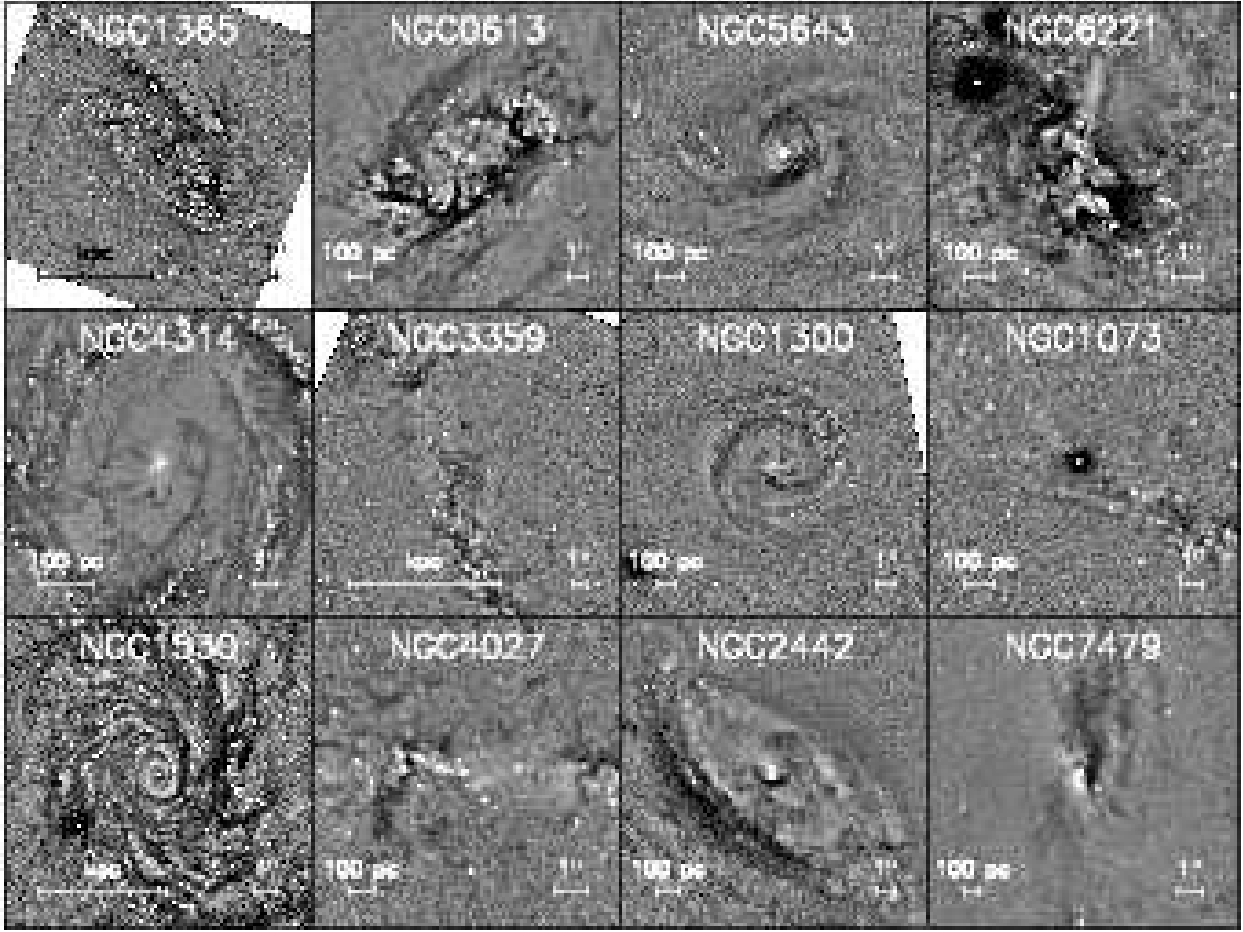}
\caption{Structure maps for the 12 most strongly barred galaxies in our
  sample ($Q_b \ge 0.4$).  $Q_b$ increases to the right and then down.
  These galaxies have LGD structure (NGC1300, NGC1365, NGC5643, NGC4314,
  and NGC1530), chaotic nuclear structure (NGC6221, NGC3359, NGC1073,
  NGC4027, and NGC7479), or a chaotic nuclear spiral (NGC0613 and
  NGC2442).  Two of the LGD spirals are associated with a circumnuclear
  ring (NGC1300 and NGC4314).  Dark regions are due to dust; bright
  regions are due to emission.
\label{fig:strong}}
\end{figure}

There remain many pieces needed to assemble the circumnuclear starburst
ring puzzle.  A large-scale bar drives dust and gas towards the center
of the galaxy.  At the boundary of the bulge and the disk, a starforming
ring develops.  Within this ring, there is a fairly coherent loosely
wound dust spiral.  It is unclear, however, whether this nuclear dust
spiral is composed of dust and gas originating from the large-scale bar
and/or the ring itself, or if it is native dust that is ``stirred-up''
by the starburst ring.  Simulations by \citet{regan04} and
\citet{maciejewski04b} imply that bars in low velocity-dispersion
systems are only capable of driving gas down to the radius at which the
ring forms---but not further---suggesting that native dust comprises the
nuclear spiral.  A galaxy with a high central velocity-dispersion,
however, could potentially have mass transported into the ring
\citep{maciejewski04b}.
\clearpage
\clearpage
For the entire sample of 75 galaxies, we find an increase in the
structure map rms $\sigma_{\mbox{\scriptsize sm}}$ with the barstrength
$Q_b$, although with substantial scatter and evidence the relation
appears to turn over at large $Q_b$.  The observed increase in dust
structure may be due to increased star formation rates, which are
correlated with the gas surface density via the global Schmidt law
\citep{schmidt59, kennicutt98}.  The structure maps for the twelve most
strongly barred galaxies in our sample ($Q_b \ge 0.4$) are shown in
Figure \ref{fig:strong}.  Of these twelve, five have LGD spirals, two of
which end at a circumnuclear ring.  The other seven have either chaotic
or chaotic spiral nuclear dust morphology (5 C, 2 CS).  As a
circumnuclear ring is associated with nuclear star formation, and the
chaotic structures seen in these strongly barred galaxies could
potentially be related to star formation, it would be interesting to
investigate the relationship between nuclear star formation rates,
circumnuclear dust morphology, dust structure $\sigma_{\mbox{\scriptsize
sm}}$, and barstrength.
\clearpage
\clearpage

\subsection{Large-scale Morphology: SB Galaxies}\label{sec:SB}
One way in which barred spiral galaxies are commonly divided into
subclasses is by taking note of where the large scale spiral starts with
relation to the bar \citep[e.g.,][]{sandage94}.  In SB(s) galaxies, the
spiral arms begin at the ends of the bar, whereas in SB(r) galaxies, the
spiral arms begin on a ring connecting the ends of the bar.  SB(rs)
galaxies are a transition group.  SB(s) structures are thought to be
preferentially found in less strongly barred galaxies than their SB(r)
counterparts \citep[e.g.,][]{sanders80, simkin80}.  Furthermore, SB(s)
galaxies typically show large-scale dust lanes that are not present in
SB(r) galaxies, and SB(r) galaxies are observed to have less dust in
their central regions than SB(s) galaxies \citep{kormendy04}.  This is
an apparent inconsistency with the generic bar-fueling picture: SB(r)
galaxies are thought to have less dust---but be more strongly
barred---while more strongly barred galaxies should have a higher
central dust content.  As these conclusions have been based on measuring
the barstrength as an axis ratio \citep[e.g.,][]{sanders80}, rather than
as a force ratio, this discrepancy might be because SB(r) galaxies are
actually more weakly barred than SB(s) galaxies.  We reinvestigate the
relation between these bar sub-types and barstrength.

The breakdown by nuclear class, including LGD, of the 21 RC3-classified
SB galaxies in our sample is given in Table~\ref{tbl:SB}.  A small
increase in $Q_b$ is seen from SB(r) to SB(s). Visually, the SB(r)
galaxies have much less dust structure than the SB(rs) and SB(s)
galaxies.  According to the Wilcoxon test, the SB(r) sample has a
smaller $Q_b$ than the SB(s) galaxies at a confidence level of 94\%.
\citet{sanders80} suggest bar pattern speed as an alternative origin to
the differences in SB(r) and SB(s) structure: a slowly rotating bar
should give rise to SB(r) structure, while a rapidly rotating bar should
yield SB(s) structure.  As the differences in barstrength are reversed
from what was expected, rotation may be more important in determining
the large-scale morphology. With respect to the amount of dust
structure, we find with $\ge 99$\% confidence that the SB(r) galaxies
have less dust structure (smaller $\sigma_{\mbox{\scriptsize sm}}$) than
the SB(s) sample; similarly, we find with 97\% confidence that the SB(r)
sample has less dust than the SB(rs) galaxies.

\clearpage
\begin{table}[htbp]
\begin{center}
\begin{tabular}{lcccc}
\hline
         &   SB    &   SB(r)  &   SB(rs)   &  SB(s) \\ \hline\hline
$\langle Q_b \rangle$  & 0.397   & 0.272    &  0.398     &    0.450 \\
Total    &   21    &   3      &   11       &  7 \\
GD       &   1     &   1      &   0        &  0 \\
TW       &   1     &   0      &   1        &  0 \\
LW       &   7     &   0      &   5        &  2 \\
CS       &   3     &   0      &   2        &  1 \\
C        &   8     &   1      &   3        &  4 \\
N        &   1     &   1      &   0        &  0 \\
LGD      &   6     &   0      &   6        &  2 \\ \hline
\end{tabular}
\caption{Average barstrengths and nuclear classifications for SB
  galaxies.  Nuclear classes are defined in \S\ref{sec:nuc}, and LGD
  morphology is defined in \S\ref{sec:gd}. SB classification is given by
  the RC3 catalog.
\label{tbl:SB}}
\end{center}
\end{table}
\clearpage

\begin{figure}
\plotone{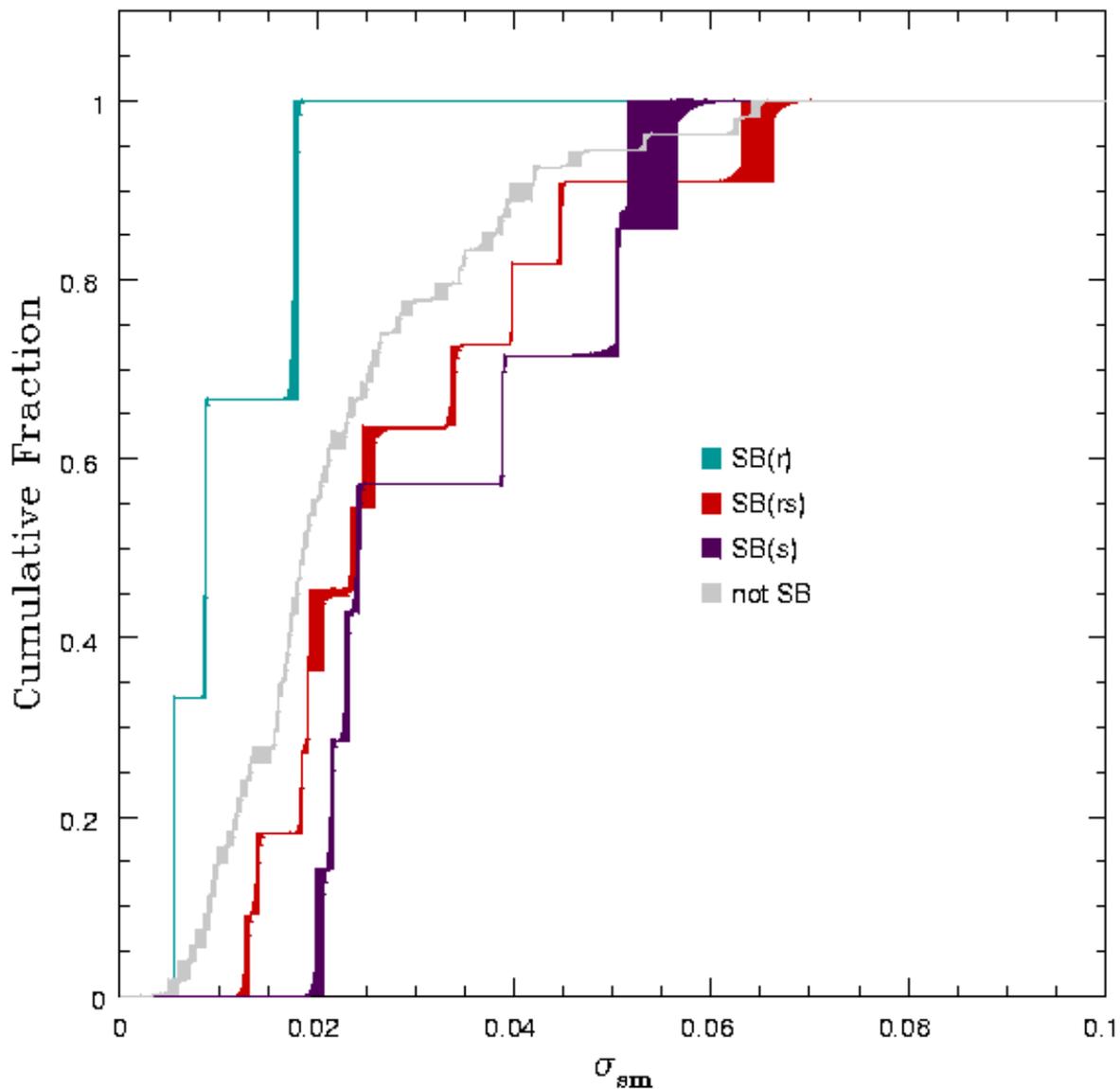}
\caption{Comparisons of structure map rms $\sigma_{\mbox{\scriptsize
sm}}$ of SB galaxies.  The vertical spreads correspond to the central
68-percentile spread of the cumulative fraction and are due to the
uncertainty in $Q_b$.  SB(r) galaxies have less dust structure than
SB(rs) galaxies, which, in turn, have less dust than SB(s) galaxies.
Specifically, with $\ge 99$\% confidence, SB(r) galaxies have less
central dust structure than SB(s) galaxies.
\label{fig:SBrms}}
\end{figure}
\clearpage
An important caveat for this aspect of our work is that large-scale
morphological classifications are known to depend on wavelength.  For
example, while only about one third of galaxies are classified as SB at
visible wavelengths, over two thirds are classified as SB in
the near infrared \citep{eskridge00}.  This is primarily because near
infrared light traces the underlying stellar mass, which is also one of
the main reasons $Q_b$ is calculated using near infrared data
\citep{buta01}.  \citet{eskridge02} reclassify 205 spiral galaxies
according to their $H$-band photometry from the OSUBGS.  Forty-eight of
our 75 galaxies are in this sample; 58\% of which (28 of the 48) are
classified as SB.  \citeauthor{eskridge02}\ do not, however, subclassify the
$H$-band images into SB(r), (rs), and (s).  It would be interesting to
see how the dust content and $Q_b$ vary by class in NIR-identified SB
galaxies.

\section{Conclusion}\label{sec:conc}
We present a study of the circumnuclear dust morphology for a sample of
75 galaxies with archival HST data and measured barstrength $Q_b$.  This
$Q_b$ is a measure of the maximal force ratio due to the presence of a
bar in a galaxy, and thus is arguably superior to the rudimentary bar axis
ratio \citep{buta01}.  We use the structure map technique of
\citet{pogge02} to enhance the visibility of the galaxies' dust content
and to classify the circumnuclear dust morphology within the central 5\%
of $D_{25}$, according to a refined version of the classification scheme
proposed by \citet{martini03a}.  We also introduce the structure map rms
$\sigma_{\mbox{\scriptsize sm}}$ within the central regions of the
galaxy as a quantitative measure of the amount of nuclear dust
structure.  A comparison of the morphological classifications and
measured barstrengths reveals that tightly wound nuclear dust spirals
(with pitch angles less than 10\dg) are preferentially found in galaxies
with lower $Q_b$.  No other nuclear class exhibits a significant
correlation with barstrength.

Previous observations found that grand design nuclear dust spirals are
hosted exclusively by strongly barred galaxies.  While we do see grand
design structure in many barred galaxies, this grand design structure
does not extend all the way into the unresolved nucleus ($\sim 10$~pc),
although it is often present at larger scales.  Earlier studies did not
strictly require that the grand design structure extend into the
nucleus.  Taking this into account, we identify two distinct types of
circumnuclear grand design spirals.  Small grand design (SGD) spirals
are nuclear dust spirals in which the two symmetric spiral arms are
coherent from 1\% of $D_{25}$ (typically a few hundred parsecs) to the
central tens of parsecs of the galaxy (where tracing the structure
becomes resolution-limited).  Large grand design (LGD) nuclear spirals,
on the other hand, show two prominent symmetric arms within 10\% of
$D_{25}$ (typically on the scale of a few kiloparsecs); these arms do
not necessarily extend to the center of the galaxy.  In fact, these two
types of grand design structure are nearly disjoint: the nuclear spiral
arms in only two of the twenty LGD galaxies in our sample extend into
the unresolved center of the galaxy (i.e., also displayed SGD
structure).  The LGD spirals are found in systematically more strongly
barred host galaxies.  This strongly confirms previous indications from
much smaller samples and demonstrates that the dust lanes along the
leading edges of large-scale spirals do not generally extend all the way
into the nuclear region, but instead lose coherence at the scale of
several hundred parsecs.  While SGD spirals are not found in galaxies
with a significantly different barstrength than typical galaxies, they
are found in galaxies with significantly less dust structure in the
central regions than LGD galaxies.  The reduced dust structure may
reflect a requirement for the formation of SGD morphology, or simply a
requirement for its detection.

The LGD spiral arms may not maintain coherence to the nucleus because
mass inflow due to the presence of a large-scale bar prompts star
formation, which can disrupt a nuclear grand design spiral.  In addition,
forty percent of the LGD spirals do not extend into the nuclear region
because there is a circumnuclear starburst ring.  We find that all of
the galaxies with circumnuclear starburst rings have LGD structure and
are more strongly barred than other galaxies.  Within the rings, three
fourths of the galaxies with circumnuclear rings have coherent loosely
wound spirals within the ring, while the others have less coherent
chaotic spirals.

We also find that among SB galaxies, SB(s) galaxies have more dust
structure and are more strongly barred than SB(r) galaxies.  This is
partially at odds with the prevailing view in the literature, which is
that SB(r) galaxies should be more strongly barred, although there is
consensus that SB(r) galaxies have less central dust
\citep[e.g.][]{kormendy04}.  \citet{sanders80} suggest that differences
in the bar pattern speed may also explain the large-scale morphological
differences between SB(s) and SB(r) galaxies; as our results indicate
that SB(s) galaxies are more strongly barred, pattern speed may be the
more relevant parameter.

Overall, there is agreement in the literature that more strongly barred
galaxies should---and do---have more dust and gas in their centers
\citep{kormendy04}.  We find that for the most strongly barred galaxies,
there are several possible morphologies the circumnuclear dust can take.
This may indicate that not all bars of a given strength $Q_b$ funnel
material toward the centers of galaxies with equal efficiency,
potentially due to the effects of pattern speed on bar efficiency, or
simply the fact that $Q_b$ is a one-parameter description of the bar.
In the most strongly barred galaxies, there can be an LGD spiral whose
arms do not extend to the galaxy nucleus but instead lose coherence.
Some LGD spirals end in circumnuclear starburst rings.  In the absence
of these structures, however, the nuclear dust in the most strongly
barred galaxies tends to be fairly chaotic, potentially hosting star
formation.  It would be interesting to investigate how circumnuclear
dust morphology and dust structure $\sigma_{\mbox{\scriptsize sm}}$
varies with nuclear star formation rates in strongly barred galaxies.

\acknowledgements{ We would like to thank Andy Gould, Rick Pogge, Witold
  Maciejewski, and the anonymous referee for their helpful comments.
  Support for this work was provided by NASA through grant AR-10677 from
  the Space Telescope Science Institute, which is operated by the
  Association of Universities for Research in Astronomy, Inc., under
  NASA contract NAS5-26555.  }

\end{document}